\documentclass[namedreferences]{Solarphysics}

\usepackage[hyperref,optionalrh]{spr-sola-addons} 
\usepackage{graphicx}        
\usepackage{color}           
\usepackage{breakurl}        

\outer\def\gtae {$\buildrel {\lower3pt\hbox{$>$}} \over 
{\lower2pt\hbox{$\sim$}} $}
\outer\def\ltae {$\buildrel {\lower3pt\hbox{$<$}} \over 
{\lower2pt\hbox{$\sim$}} $}

\begin{document}

\outer\def\gtae {$\buildrel {\lower3pt\hbox{$>$}} \over 
{\lower2pt\hbox{$\sim$}} $}
\outer\def\ltae {$\buildrel {\lower3pt\hbox{$<$}} \over 
{\lower2pt\hbox{$\sim$}} $}
\newcommand{\Msun}{$M_{\odot}$}
\newcommand{\lsun}{$L_{\odot}$}
\newcommand{\Rsun}{$R_{\odot}$}
\newcommand{\Solar}{${\odot}$}
\newcommand{\kep}{\sl Kepler}
\newcommand{\ktwo}{\sl K2}
\newcommand{\tess}{\sl TESS}
\newcommand{\swift}{\it Swift}
\newcommand{\Porb}{P_{\rm orb}}
\newcommand{\nuorb}{\nu_{\rm orb}}
\newcommand{\eplus}{\epsilon_+}
\newcommand{\eminus}{\epsilon_-}
\newcommand{\cd}{{\rm\ c\ d^{-1}}}
\newcommand{\MdotL}{\dot M_{\rm L1}}
\newcommand{\Mdot}{$\dot M$}
\newcommand{\MdotSolar}{\dot{M_{\odot}} yr$^{-1}$}
\newcommand{\Ldisk}{L_{\rm disk}}
\newcommand{\src}{KIC 9202990}
\newcommand{\ergscm} {erg s$^{-1}$ cm$^{-2}$}
\newcommand{\rchi}{$\chi^{2}_{\nu}$}
\newcommand{\chisq}{$\chi^{2}$}
\newcommand{\pcmsq} {cm$^{-2}$}

\begin{article}
\begin{opening}
\title{TESS observations of flares and quasi-periodic pulsations from low mass stars and potential impact on exoplanets}

\author[addressref={aff1},corref,email={gavin.ramsay@armagh.ac.uk}]{\inits{G.}\fnm{Gavin}~\lnm{Ramsay}}
\author[addressref={aff2,aff3},email={D.Kolotkov.1@warwick.ac.uk}]{\inits{D.}\fnm{Dmitrii}~\lnm{Kolotkov}}
\author[addressref={aff1},email={gerry.doyle@armagh.ac.uk}]{\inits{J.G.}\fnm{J.Gerry}~\lnm{Doyle}}
\author[addressref={aff4,aff5},email={lauren.doyle@warwick.ac.uk}]{\inits{L.}\fnm{Lauren}~\lnm{Doyle}}
\address[id=aff1]{Armagh Observatory and Planetarium, College Hill, Armagh, BT61 9DG, N. Ireland, UK}
\address[id=aff2]{Centre for Fusion, Space and Astrophysics, Department of Physics, University of Warwick, Coventry, CV4 7AL, UK}
\address[id=aff3]{Institute of Solar-Terrestrial Physics SB RAS, Irkutsk 664033, Russia}
\address[id=aff4]{Centre for Exoplanets and Habitability, University of Warwick, Coventry, CV4 7AL, UK}
\address[id=aff5]{Department of Physics, University of Warwick, Coventry, CV4 7AL, UK}

\runningauthor{Ramsay et al.}
\runningtitle{QPPs from low mass stars}

\begin{abstract}
We have performed a search for flares and Quasi-Periodic Pulsations
(QPPs) from low mass M dwarf stars using TESS 2 min cadence data. We
find seven stars which show evidence of QPPs. Using Fourier and
Empirical Mode Decomposition techniques, we confirm the presence of 11
QPPs in these seven stars with a period between 10.2 and 71.9 min,
including an oscillation with strong drift in the period and a
double-mode oscillation. The fraction of flares we examined which
showed QPPs (7 percent) is higher than other studies of stellar
flares, but is very similar to the fraction of Solar C-class
flares. Based on the stellar parameters taken from the TESS Input
Catalog, we determine the lengths and magnetic field strengths of the
flare coronal loops using the period of the QPPs and various
assumptions about the origin of the QPPs. We also use a scaling
relationship based on flares from Solar and Solar-type stars and the
observed energy, plus the duration of the flares, finding that the
different approaches predict loop lengths which are consistent to a
factor of $\sim$2. We also discuss the flare frequency of the seven
stars determining whether this could result in ozone depletion or
abiogenesis. Three of our stars have a sufficiently high rate of
energetic flares which are likely to cause abiogenesis. However, two
of them are also in the range where ozone depletion is likely to
occur. We speculate on the implications for surface life on these
stars and the effects of the loop lengths and QPPs on potential
exoplanets in the habitable zone.
\end{abstract}
\end{opening}

\section{Introduction}

The Sun’s variable magnetic activity influences its surrounding
heliosphere which leads to a variety of observed phenomenon from
small-scale features such as spicules to large-scale events such as
flares, coronal mass ejections; the latter of these leads to space
weather which affects Earth. Flares and coronal mass ejections pose a
danger to electric power grids and telecommunications facilities,
satellites and astronauts, e.g.
\citep{NationalResearchCouncil2008}. Furthermore, the Sun’s radiative
output can affect planetary and global climate on much longer
timescales from decades to stellar evolutionary timescales,
e.g. \citep{Mursula2007,Nandy2021}. However, we now know that the Sun
is much less active than most Solar-type stars, although it remains
unclear whether the Sun was always less active or whether its activity
levels have declined over millions of years,
e.g. \citet{Reinhold2020}.

Unlike Solar type stars which have a radiative core and a convective
envelope, stars with a mass \ltae0.4\Msun (corresponding to a
$\sim$M4V or later spectral type) are fully convective. The fraction
of low mass stars which show optical flares increases from M1V
($\sim$5 percent) to M6V \citep[$\sim$45 percent,][]{Gunther2020}
showing low mass stars are active.  One of the principal factors in
determining the degree of flare activity is a stars age, with activity
declining as stars get older, e.g. \citet{Skumanich86} and more
recently by \citet{Davenport2019}. Understanding stellar activity in
general has become an area of renewed interest for several
reasons. Stellar activity can mask or give false positive detections
of exoplanets, e.g.  \citet{Rajpaul2015} and stellar flares can effect
the atmosphere of planets orbiting their host star, e.g.
\citet{Airapetian2020}. However, in more recent years, it has been
argued that the UV flux incident on an exoplanet, which flares can
deliver, are essential for life to form, e.g.  \citet{Rimmer2018}.

In recent years, there has been major advancements in the detection
and analysis of quasi-periodic pulsations (QPPs) in Solar and stellar
flares. These QPPs can appear at all phases of a flare from the
impulsive to decay phase. Based on a number of statistical studies,
QPPs are shown to be a frequent and wide-spread phenomena. There are
over a dozen possible mechanisms which produce oscillations in a
plasma. In flares from M dwarfs, we have seen sub-second pulses in
radio bursts \citep{Osten2006,Osten2008}, a few tens of seconds in
ultra-violet data \citep{Doyle2018} and tens of minutes in optical
data from {\sl Kepler} \citep{Pugh2016}. The proposed models include
magnetohydrodynamic waves, repetitive reconnection and oscillations in
current sheets. It is also possible that different mechanisms operate
in different flares. By studying these properties we can gain
important insights to the physical nature of flares and their
immediate environment. This allows for the development of theoretical
models which explain the origin and properties of both Solar and
stellar flares.

The first detection of QPPs in stellar flares was from a M4e star
which showed oscillations in the optical on a period of around a dozen
seconds using photoelectric observations \citep{Rodono1974}. It was
much later that QPPs were also seen in X-ray observations of a stellar
flare, this time on a period of a dozen minutes
\citep{MitraKraev2005}. QPPs with timescales shorter than a dozen
minutes have now been seen from many stars, see also
\citet{Balona2015}. In addition to those given by \citet{Pugh2016}, an
example of a long period QPP was from YZ CMi (M4.5e) which had a
period of 32 minutes \citep{Anfinogentov2013}.  \citet{Reale2018}
report X-ray observations of 3 hr pulsations in two pre-main sequence
stars, implying a very large stellar loop structure.  \citet{Cho2016}
made a comparison between the observed characteristics of Solar and
stellar QPPs seen in X-rays and concluded that the underlying
mechanism responsible was the same in both the Solar and stellar
atmospheres. For reviews of QPPs from Solar and stellar flares see
\citet{McLaughlin2018,VanDoorsselaere2016,Kupriyanova2020,Zimovets2021}.

The means of detecting flares from many stars simultaneously has been
transformed with the {\kep} and {\tess} missions. {\kep} stared at the
same 115 square degree field of view for nearly four years resulting
in hundreds of flares being observed from stars of different spectral
type \citep[e.g.][]{Davenport2016}. {\tess} has now observed a
  large fraction of the sky with photometry available for each sector
of sky, each lasting approximately a month in duration. Flares have
been seen from stars including Solar type stars
\citep[e.g.][]{Doyle2020,Tu2020} and M dwarfs
\citep[e.g.][]{Ramsay2020,Gunther2020}. Observations of the nearest
star to our Sun, Proxima Centauri (M5.5V), made using {\tess}, showed
two flares with QPPs on a timescale of a few hrs \citep{Vida2019}
indicating that {\tess} could open up a large sample of QPP events
from low mass stars.

In this paper we use {\tess} data taken with 2 min cadence to search
for high amplitude flares from low mass stars. We identify those which
have relatively long duration events (a few hrs) and show evidence for
QPPs in the decline from maximum. We apply a sophisticated set of
tests to determine the significance of the candidate QPPs and then
determine the length of the flare loop structures based on the stars
radius and mass. We also discuss the effects of the high energy flares
on the atmosphere of exoplanets and whether QPPs themselves could make
an impact. Finally we draw parallels between stellar and Solar
activity.

\section{{\tess} observations}
\label{tess}

{\tess} was launched in April 2018 and consists of four 10.5 cm
telescopes that observe a 24$^{\circ}\times96^{\circ}$ strip (known as
a {\it sector}) of sky for $\sim$28~days \citep[see][for
  details]{Ricker2015}. Between July 2018 and June 2019, {\tess}
covered most of the southern ecliptic hemisphere (Cycle 1) and between
July 2019 and June 2020 covered most of the northern ecliptic
hemisphere (Cycle 2). Although there is a band along the ecliptic
plane which was not observed, at the ecliptic poles there is a
continuous viewing zone where stars can be observed for $\sim$1~yr
(to avoid stray light from the Earth and Moon some areas of the
  northern hemisphere were not observed as originally intended). Each
`full-frame image' has an exposure time of 30 min. However, in each
sector, photometry with a cadence of 2 min is obtained, with most
targets being selected from the community via a call for proposals.

Since our study requires well sampled light curves with enough
resolution to resolve QPPs, we have used 2 min cadence data from
Cycles 1 and 2. Using the stars observed in each sector, we find that
217,834 unique stars were observed in 2 min cadence in Cycles 1 and 2
which were also in Gaia DR2 \citep{Gaia2018}. To select low mass stars
we cross-matched the sky co-ordinates of our sample with that of Gaia
DR2 and obtained the stars $(BP-RP)$ colour. We derived their absolute
$G$ mag using the parallax and a Galactic scale length L=1.35~kpc
\citep{Astraatmadja2016}. Following on from \citet{Ramsay2020} we
chose to select stars which have a Gaia colour $(BP-RP)>1.8$ and are
close to the main sequence. Our selection excludes stars which are
likely to be in binary systems; are likely younger than
$\sim$30~Myr or have significant reddening (which is unlikely for
relatively bright nearby M dwarfs): this gave 15,437 unique stars.

We downloaded the calibrated light curves of our targets from the MAST
data archive\footnote{\url{https://archive.stsci.edu/tess/}}. We
initially used the data values for {\tt SAP\_FLUX}, which are the
Simple Aperture Photometry values derived using the standard pipeline
and normalised each light curve to unity on a sector by sector
basis. For those stars with data from more than one sector we made one
combined light curve.

For each of the 15,437 light curves, we identified those which had at
least one point with a maximum of 1.6 times the mean (corresponding to
0.5 mag in amplitude) and was also 5 times the rms of the full light
curve. Each light curve which passed this selection procedure was
manually inspected to ensure the event was not due to obvious
instrumental effects or a minor planet passing over the field, which
show a symmetric profile.

A total of 481 stars showed at least one flare with an amplitude of
0.5 mag and also passed our initial manual verification phase. Of this
sample, 178 stars had at least one high amplitude flare which had a
duration greater than 1~hr, with 40 having a duration of 2~hr or
more. The light curve of each of the 178 stars were then visually
inspected to search for candidate QPPs in the decline from flare
maximum. We identified seven stars which show evidence for candidate
QPPs events: three stars showed more than one candidate QPP event in
separate flares.

\begin{figure}
  \begin{center}
  \includegraphics[width=0.87\textwidth]{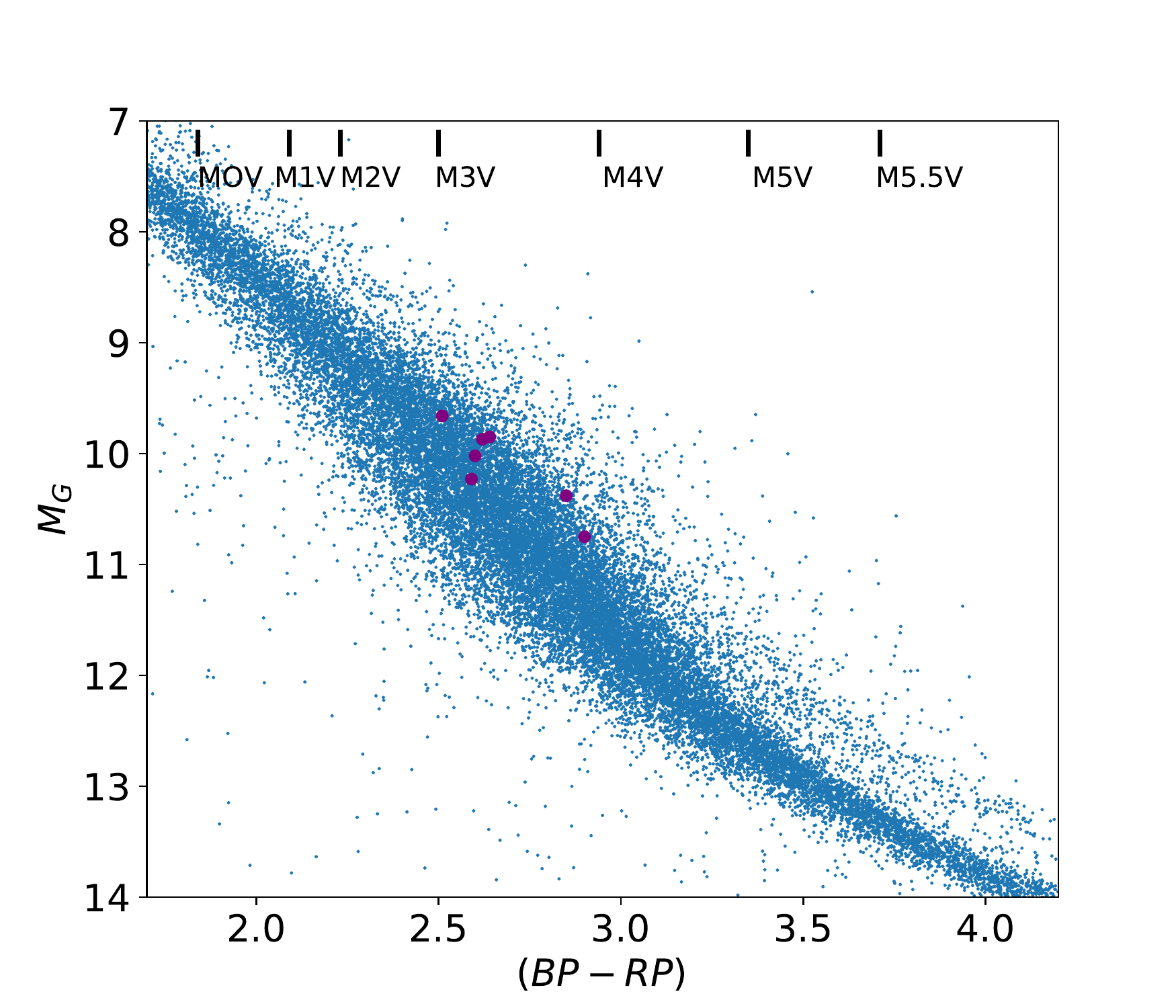}    
\vspace{2mm}
  \caption{Stars within 50 pc are shown as small blue dots in the Gaia
    $(BP-RP)$, $M_{G}$ plane. Purple dots show those stars which we
    identified as having candidate QPPs in at least one flare
    event. Using the work of \citet{PecautMamjek2013}, we find they
    have a spectral type in the range M3--M4V.}
    \label{gaiaHRD}
    \end{center}
\end{figure}

The physical and observational characteristics of these stars are
outlined in Table \ref{stellar_properties}. We show in Fig.
\ref{gaiaHRD} the position of these stars in the Gaia $(BP-RP)$,
$M_{G}$ plane. This indicates that they have spectral types around
M3--M4V\footnote{\url{https://www.pas.rochester.edu/~emamajek/EEM_dwarf_UBVIJHK_colors_Teff.txt}},
the point where stars become fully convective.

\begin{figure}
  \centering
	\includegraphics[width=0.85\linewidth]{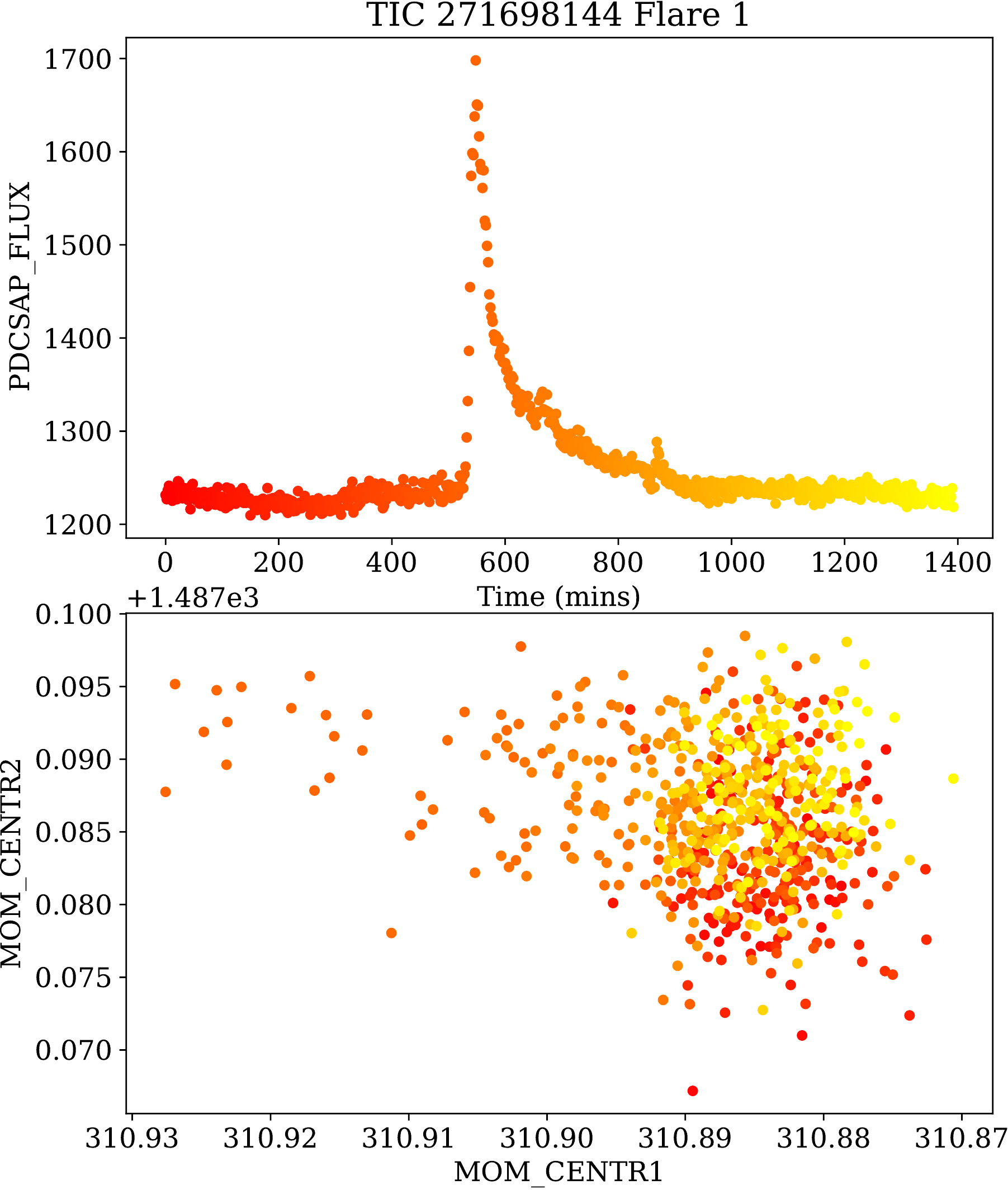}
	\includegraphics[width=0.85\linewidth]{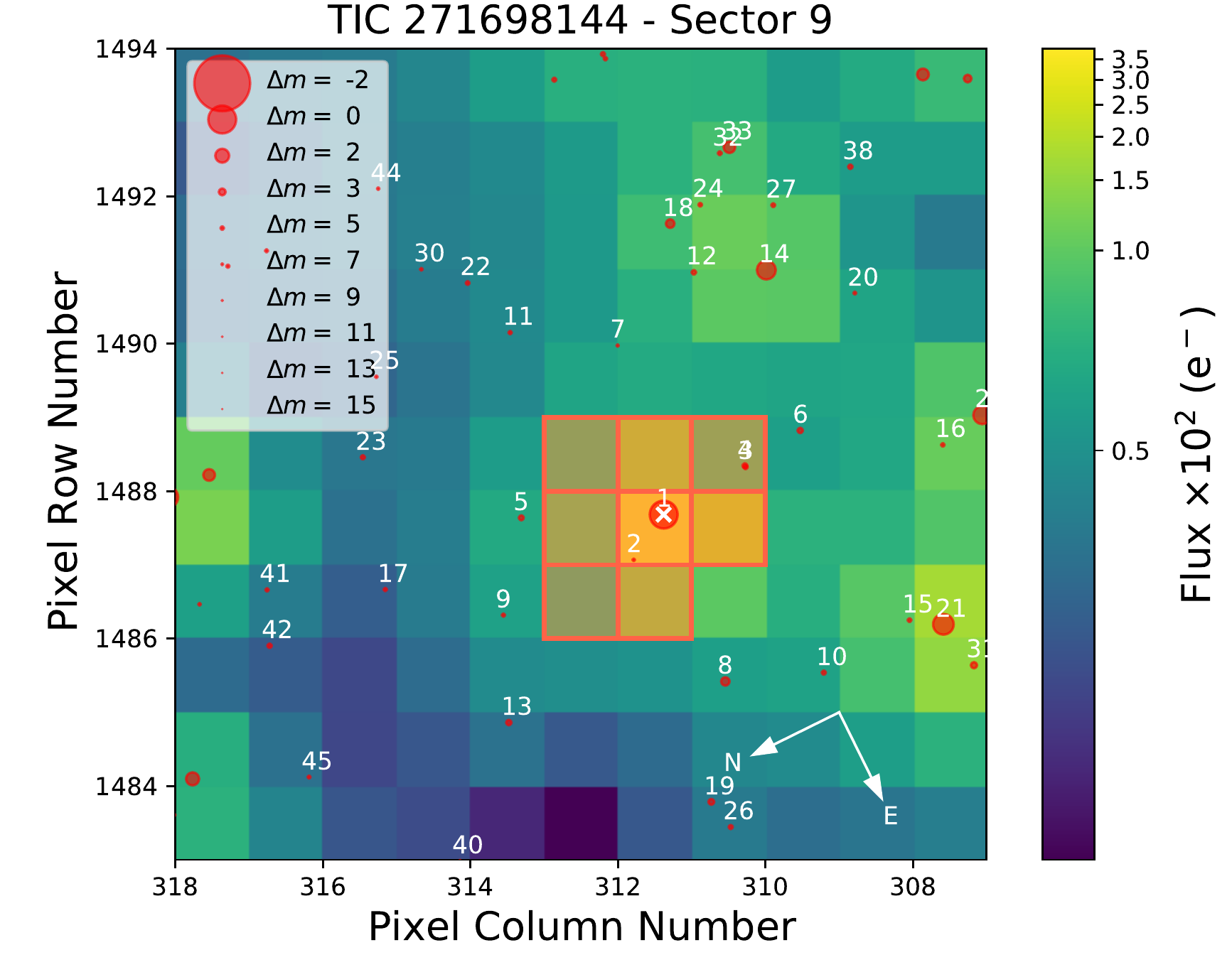}
\caption{In the top panel we show the light curve of the first flare
  we detect in TIC 271698144 colour coded in such a way that it
  reflects the detector coordinates in the middle panel. Middle panel
  indicates that during the flare the centroid of the photocenter
  shifts to the upper left. The lower panel shows a map of the
  immediate field taken from the Target Pixel File with the aperture
  map shown as red boxes. The spatially nearby stars are all at least
  5 mag fainter than the target. During the flare, the photocenter
  shifts {\sl towards} the target. We attribute this to chromatic
  aberration.}
\label{xyGaia}
\end{figure}

We then considered the issue of the large pixel size (21 arcsec per
pixel) of the detectors on the {\tess} cameras which can cause
variability from spatially nearby stars to contaminate the light curve
of the target star. We initially used {\tt tpfplotter}
\citep{Aller2020} to overlay the position of stars in the Gaia DR2
catalogue onto an image derived from a TESS Target Pixel File (see the
lower panel of Fig. \ref{xyGaia}). This indicates which, if
any, stars were in the aperture mask used to extract the light curve
of the target, which typically is 7--8 pixels in size. Of the seven
stars in our sample, three had no stars within 3 mag of the target in
the aperture mask. For the remaining four stars, the stars were
between 2.8 -- 2.9 mag fainter than the target. Using Gaia EDR3 data
to determine their place on the Gaia HRD we identify the spatially
nearby stars as being late K or early M spectral types.

The Target Pixel files also allowed us to examine how the position
of the center of the photocentre of the Point Spread Function (PSF)
({\tt MOM$_{CENTR1/2}$}) varied over the observation. It is quite
normal for the position of the photocenter to shift by a few 0.01
pixels in x,y co-ordinates over the $\sim$13--14 days which make up
half of a sector's observations. What we find is that even for stars
which have no spatially nearby stars in the aperture mask, during
more intense flares the photocenter can shift by $\sim$0.1 pixels in
x,y co-ordinates. As an example, we show in the top panel of 
Fig. \ref{xyGaia} the position of the centroid of the
photocenter of TIC 271698144 (one of our selected objects) over the
course of a high amplitude flare: it shifts in x,y by a few 1/100 of
a pixel {\sl towards} the target (although in other cases it
  was not). We also show a map of the immediate field with the
location of the aperture mask shown. The spatially nearby stars are
at least 5 mag fainter than the target.  The only explanation we can
attribute to this observation is chromatic aberration which
  will be sensitive to position of the target in the plane of the
  detector. During a high amplitude flare, the colour of the
incident light is hotter (bluer) and therefore could in principal be
slightly offset from the previous centroid. This has been studied in
depth for {\sl Kepler} data by \citet{Hedges2021} who show the shape
of the PSF varies depending on the colour and note that due to the
refractive nature of the {\tess} optics this effect is likely to be
much greater in {\tess} data. Finally, we note that
\citet{Jackman2021} made a systematic study of flares from M dwarfs
using {\kep} and {\tess} data and conclude that for {\tess} data
there is a 5.8 percent chance of a false positive flare event due to
spatially nearby stars. We conclude that it is highly unlikely that
the flares in which we detect candidate QPPs do not originate from
the M dwarf target.

\begin{table}
\caption{The stellar parameters for the sources which we have
  identified QPPs in at least\\
  one flare observed using {\tess} 2 min
  cadence data. The data has been extracted from the\\
  TIC v8.0 \citep{Stassun2019} and Gaia DR2.}
\begin{center}
\resizebox{\textwidth}{!}{
\begin{tabular}{rrrrrrrrrrrrr}
\hline
  \multicolumn{1}{c}{TICID} &
  \multicolumn{1}{c}{Tmag} &
  \multicolumn{1}{c}{RA} &
  \multicolumn{1}{c}{DEC} &
  \multicolumn{1}{c}{pmra} &
  \multicolumn{1}{c}{pmdec} &
  \multicolumn{1}{c}{$BP-RP$} &
  \multicolumn{1}{c}{D} &
  \multicolumn{1}{c}{MG} &
  \multicolumn{1}{c}{L} &
  \multicolumn{1}{c}{$T_{\rm{eff}}$} &
  \multicolumn{1}{c}{Radius} &
  \multicolumn{1}{c}{Mass} \\
  \multicolumn{1}{c}{} &
  \multicolumn{1}{c}{} &
  \multicolumn{1}{c}{(2000)} &
  \multicolumn{1}{c}{(2000)} &
  \multicolumn{1}{c}{(mas/yr)} &
  \multicolumn{1}{c}{(mas/yr)} &
  \multicolumn{1}{c}{} &
  \multicolumn{1}{c}{(pc)} &
  \multicolumn{1}{c}{} &
  \multicolumn{1}{c}{(erg/s)} &
  \multicolumn{1}{c}{(K)} &
  \multicolumn{1}{c}{(\Rsun)} &
  \multicolumn{1}{c}{(\Msun)} \\
  \hline
  1403938 & 13.84 & 140.26158 & -15.89001 & -92.3$\pm$0.1 & 30.7$\pm$0.1 &  2.85 & 83.9$\pm$0.8 &  10.38 & 5.0E31 & 3251$\pm$157 & 0.371$\pm$0.011 & 0.358$\pm$0.02\\
  233547261 & 13.24 & 281.73833 & 60.89595 & 14.8$\pm$0.1 &  -26.5$\pm$0.1 & 2.60 & 78.1$\pm$0.4  & 10.02 & 6.4E31 & 3376$\pm$157 & 0.390$\pm$0.012 & 0.381$\pm$0.02\\
  279494336 & 12.9 & 59.08738 & -15.80579 & -13.7$\pm$0.1 & -46.7$\pm$0.1 & 2.62 & 72.9$\pm$0.5 &  9.87 & 6.9E31 & 3366$\pm$157 & 0.407$\pm$0.012 & 0.399$\pm$0.02\\
  353898013 & 12.63 & 271.89985 & 56.32422 & 22.9$\pm$0.1 & 41.2$\pm$0.1  & 2.90 & 43.9$\pm$0.2 & 10.75 & 3.8E31 & 3230$\pm$157 & 0.328$\pm$0.010 & 0.309$\pm$0.02\\
  455825451 & 12.19 & 322.41897 & 64.09441 & 90.9$\pm$0.1 & 28.8$\pm$0.1 & 2.59 & 43.8$\pm$0.1 & 10.23 & 5.8E31 & 3384$\pm$157 & 0.369$\pm$0.011 & 0.356$\pm$0.02\\
  271698144 & 12.63 & 110.13863 & -75.76763 & -35.8$\pm$0.1 &71.4$\pm$0.1 & 2.51  & 69.6$\pm$0.3 & 9.66 & 8.3E31 & 3429$\pm$157 & 0.431$\pm$0.013 & 0.426$\pm$0.02\\
  393804343 & 12.93 & 81.32051  & -20.90736 & 20.8$\pm$0.1  &-40.7$\pm$0.1 & 2.64 & 74.8$\pm$0.6 & 9.85 & 7.2E31 & 3361$\pm$157 & 0.418$\pm$0.012 & 0.411$\pm$0.02\\
  \hline
  \end{tabular}}
  \end{center}
  \label{stellar_properties}
\end{table}

\begin{table}
\begin{center}
\begin{tabular}{lrrrrrr}
\hline
TIC & Duration & Period & Amp    & Max        & No. & No. \\
    &          &        &        & Energy     & Flares & Flares\\
    & ( d)     & (d)    & (frac) & (erg)      &  & with QPPs\\
\hline
1403938   &  18.6 & 0.645 & 0.0104 & 6.0$\times10^{35}$ & 2 & 2\\
233547261 & 321.6 & 2.45  & 0.0380 & 2.3$\times10^{36}$ & 24 & 1\\
279494336 &  46.9 & 0.924 & 0.0060 & 6.1$\times10^{35}$ & 2 & 1\\
353898013 & 321.5 & 1.659 & 0.0294 & 1.1$\times10^{36}$ & 79 & 1\\
455825451 &  95.5 & 0.478 & 0.0332 & 7.8$\times10^{35}$ & 31 & 1\\
271698144 & 287.0 & 0.419 & 0.0100 & 8.1$\times10^{35}$ & 85 & 2\\
393804343 &  45.2 & 0.912 & 0.0202 & 8.7$\times10^{35}$ & 17 & 3\\
\hline
\end{tabular}
\label{periodetc}
\caption{For those stars showing candidate QPPs, we show the duration
  of the {\tess} light curve; the period which we take to be the
  rotation period; the amplitude of the rotation period; the maximum
  bolometric energy of the flares seen, the total number of flares
  detected and the number of flares with QPPs.}
\end{center}
\end{table}

For each of the seven stars which passed this verification phase we
took the bolometric luminosity of our target stars from the TIC V8.0
catalogue \citep{Stassun2019}. Determining the luminosity of the
flares involves some degree of approximation. Unlike the photosphere
of the M dwarf star, the temperature of the flare can be considerably
higher. We assume that the temperature of the flare is $\sim$12,000~K
and that the fraction of the emitted flux which falls within the
{\tess} pass-band is $\sim$0.14 \citep{Schmitt2019}. This implies a
correction factor of $\sim$7 to obtain the bolometric luminosity of
the flare (this gives flare energies $\sim$1.5 greater
  compared with a temperature of 9,000K, \citep{Howard2019}). With this
  in place, we then removed the signature of the rotational modulation
  and instrumental effects using a routine in the {\tt lightkurve}
  python package \citep{lightkurve2018}. We then searched for flares and
  calculated the energy of {\sl all} the flares in that light curve
  using the {\tt
    Altaipony}\footnote{\url{https://altaipony.readthedocs.io/en/latest}}
  suite of python based software which is an update of the {\tt
    Appaloosa} \citep{Davenport2016} suite of software. To determine
the likely rotation period of the stars, we used the light curves
  derived from the {\tt PDCSAP\_FLUX} values, which are the Simple
  Aperture Photometry values, {\tt SAP\_FLUX}, after correction for
  systematic trends and the Lomb Scargle (LS) Periodogram as
implemented in the {\tt VARTOOLS} suite of tools
\citep{HartmannBakos2016}. We show the periods of each star in Table
\ref{periodetc}: where (i) each period was clearly seen in the light
curve, (ii) has a very low False Alarm Probability and (iii) is within
the range 0.4--2.5~d. We now go on to a detailed assessment of the
candidate QPP events, returning to the wider implications of the rate
of high energy flares from these stars in \S \ref{flarerates}.

\begin{figure*}
	\centering
	\includegraphics[width=\linewidth]{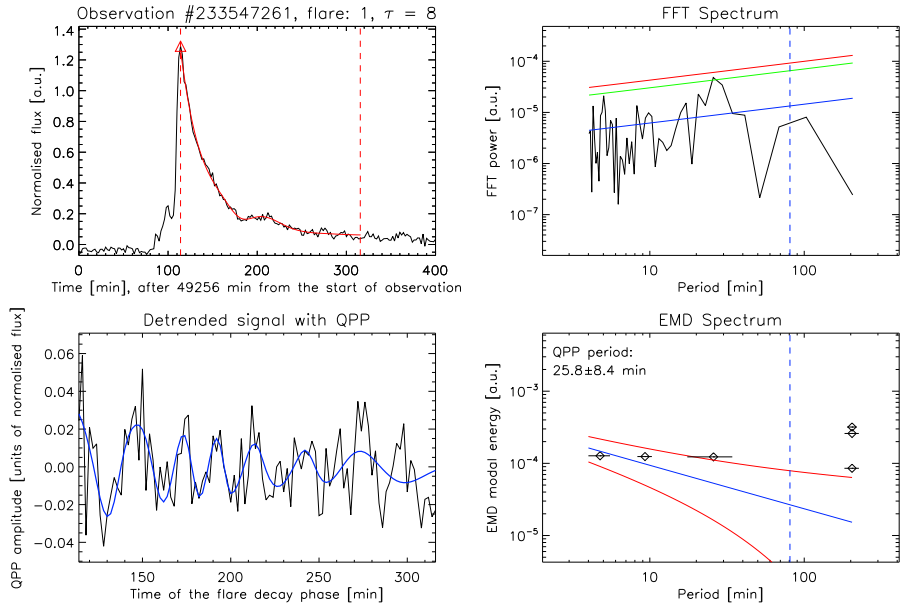}
	\caption{A QPP event manifesting strong drift in the
          oscillation period.  Top left: the flare light curve (the
          black line) and a low-frequency trend $T(t)$ of its decay
          phase (the red line), obtained as described in
          Sec.~\ref{sec:search_qpp}, step \ref{item:trend}. The
          vertical dashed lines indicate the time interval of
          interest. The red triangle shows the position of the
          apparent flare peak (the highest value in the light curve).
          Bottom left: The original flare section of interest with the
          low-frequency trend subtracted (the black line). The blue
          line shows the EMD-revealed statistically significant
          intrinsic mode.  Top right: the Fourier power spectrum of
          the detrended flare signal shown in the bottom left panel in
          black. The blue solid line shows the best-fit of the
          spectrum by a power law function representing a
          superposition of the white and coloured noise. The green and
          red solid lines indicate the statistical significance levels
          of 1-$\sigma$ (68\%) and 2-$\sigma$ (95\%), estimated as
          described in Sec.~\ref{sec:search_qpp}, step
          \ref{item:fourier}. The vertical blue dashed line indicates
          the oscillation period equal to 0.4 of the total analysed
          signal length. All the EMD-revealed intrinsic modes with
          mean periods longer than what could be attributed to the
          low-frequency trend $T(t)$ in this work.  Bottom right: The
          EMD spectrum (dependence between the EMD-revealed modal
          energies and mean periods) of the original flare section of
          interest with the exponential trend $T_\mathrm{exp}(t)$
          subtracted, see steps \ref{item:trend_exp} and
          \ref{item:emd} in Sec.~\ref{sec:search_qpp}. The blue solid
          line shows the expected behaviour of a power-law distributed
          noise with parameters estimated from the Fourier
          analysis. The red solid lines show the confidence levels of
          95\% estimated as described in Sec.~\ref{sec:search_qpp},
          step \ref{item:emd_confidence}. The error bars for the
          values of the EMD-revealed mean modal periods are estimated
          as the half-level-width of the global wavelet spectra
          calculated for each of the intrinsic modes and best-fitted
          by the Gaussian function.  }
	\label{fig:qpp_drift}
\end{figure*}

\begin{figure*}
	\centering
	\includegraphics[width=\linewidth]{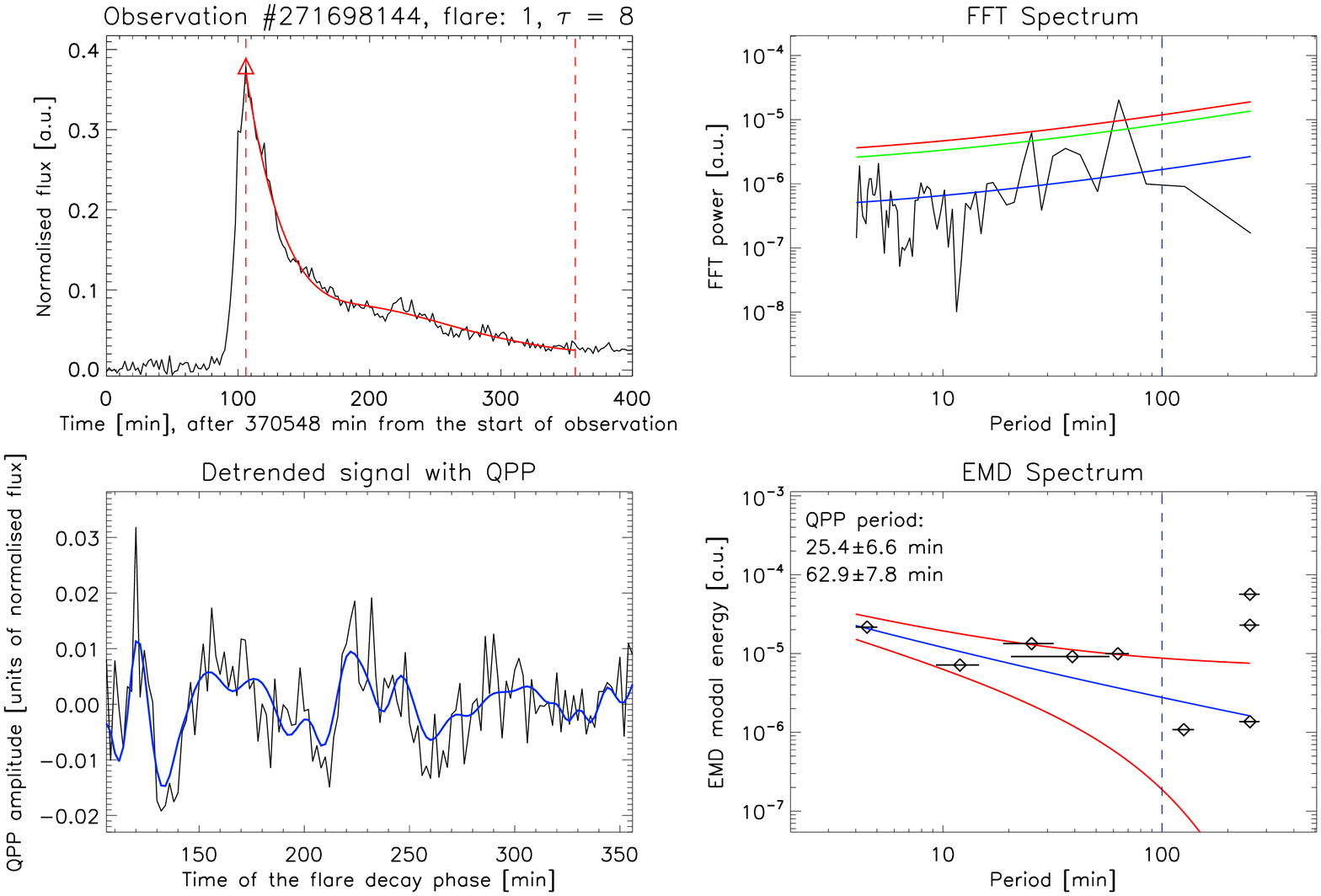}
	\includegraphics[width=\linewidth]{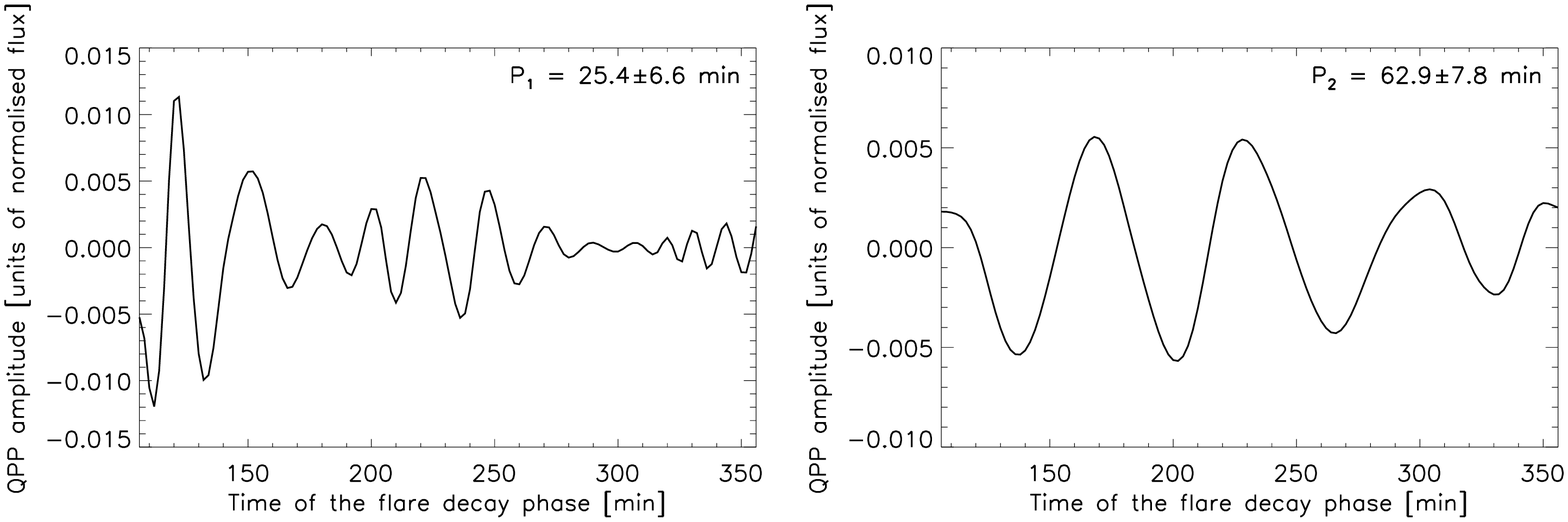}
	\caption{A two-modal QPP event. The layout and notations of
          the top and middle panels are identical to those shown in
          Fig.~\ref{fig:qpp_drift}. The two bottom panels show the
          individual EMD-revealed modes of this two-modal QPP event.
        }
	\label{fig:multi-modal_qpp}
\end{figure*}

\section{Searching for QPPs}
\label{sec:search_qpp}

We found seven low mass M dwarf stars with candidate QPPs in
their {\tess} light curves. How to detect the presence of QPPs in
stellar flares and determine their significance is a challenging task
with many different approaches having been used over the
years. \citet{Broomhall2019} outline the potential pitfalls in these
approaches and make a number of recommendations for such searches.

For identifying the flare events to be checked for the presence of QPP
patterns in all of the available observations, we begin with assuming
$t_\mathrm{rise} = 100$~min for pre-flare and flare rising phases and
$t_\mathrm{dec} = 500$~min for the flare decay phase. For each of the
light curves, we set a threshold of 3-$\sigma$, where the standard
  deviation was derived from the full light curve and identify all
local maxima exceeding this threshold. From the identified local
maxima, we select only those where the time interval is greater than
$0.3t_\mathrm{dec}$ (thus we consider the events with multiple maxima
situated closer than $0.3t_\mathrm{dec}$ with respect to each other as
a single event). The selected maxima preceded by $t_\mathrm{rise}$ and
followed by $t_\mathrm{dec}$ thus become the list of flaring events
for the further QPP analysis (see the red triangles in
Figs.~\ref{fig:qpp_drift}-\ref{fig:multi-modal_qpp} and
Figs.~\ref{fig:qpp_all_1}-\ref{fig:qpp_all_3}).

The analysis of QPP signals that we carried out in this work
represents a synergy of the methods used by \citet{Broomhall2019}: (i)
Fourier transform with detrending by smoothing and not taking the
background coloured noise into account, (ii) a Fourier transform
without detrending and accounting for the background coloured noise,
and (iii) a method of empirical mode decomposition (EMD) with a
self-consistent detrending and assessment of the statistical
significance of the revealed intrinsic oscillatory modes in comparison
with the background coloured noise. More specifically, for each of the
identified flare events,
\begin{enumerate}
	\item We focus on the decay phase of the flare, for the
          beginning of which we use the time of the flare peak
          (highest value in the flare light curve), $t_\mathrm{peak}$;
          and the above-mentioned $t_\mathrm{dec}=500$\,min for its
          duration.
	\item As a rough estimate of the flare $e$-folding time
          $t_{1/e}$, we best-fit the decay phase of the flare with an
          exponential function.\label{item:exp_fit}
	\item To account for the effects of flare trimming on the
          detection of QPPs \citep[see Sec.~5.3 in][]{Broomhall2019},
          we vary the length of the signal of interest as
          $[t_\mathrm{peak},~t_\mathrm{peak}+\tau t_{1/e}]$, where
          $\tau$ is an integer number varying from 2 to 10 in this
          work.
	\item From the resulting flare sections
          (i.e. $[t_\mathrm{peak},~t_\mathrm{peak}+2 t_{1/e}]$,
          $[t_\mathrm{peak},~t_\mathrm{peak}+3 t_{1/e}]$, etc.), we
          select only those in which the number of observational data
          points is greater than 30. Fewer data points would not allow
          for a meaningful Fourier or EMD analysis of periodicities.
	\item For each of those selected flare sections $F_0(t)$, we
          subtract the exponential fit $T_\mathrm{exp}(t)$ obtained at
          step \ref{item:exp_fit}, as a rough approximation of the
          flare trend. This gives $F_1(t)\equiv F_0(t) -
          T_\mathrm{exp}(t)$. This allows us to mitigate the
          discontinuity between the start and end points in $F_0(t)$,
          which is crucial for the Fourier and EMD
          techniques.\label{item:trend_exp}
	\item We apply the EMD method to $F_1(t)$. From all the
          EMD-revealed intrinsic modes, we combine all modes with
          characteristic time scales longer than 0.4 of the total
          signal length (so that the number of oscillation cycles in
          each of those modes is less than 2.5) in a slowly varying
          trend $T_\mathrm{emd}(t)$ of $F_1(t)$.\label{item:emd}
	\item We define the total trend of the original flare signal
          $F_0(t)$ as $T(t) \equiv T_\mathrm{exp}(t) +
          T_\mathrm{emd}(t)$ (see the red solid lines in the top left
          panel of
          Figs.~\ref{fig:qpp_drift}-\ref{fig:multi-modal_qpp}), and
          subtract $T(t)$ from $F_0(t)$ to obtain the detrended signal
          $F(t)$.\label{item:trend}
	\item We apply the Fourier analysis to $F(t)$ (see the top
          right panels in
          Figs.~\ref{fig:qpp_drift}-\ref{fig:multi-modal_qpp}) and
          assess the significance of the Fourier peaks and obtain
          parameters of the background noise (i.e. its power law index
          also known as the noise \lq\lq colour\rq\rq, and energies of
          the white and coloured components), adapting the methods
          from \citet{Vaughan2005} and \citet{Pugh2017}. We note here
          that due to the intrinsic non-stationary properties of QPP
          (i.e. short lifetime, modulation of the oscillation
          amplitude and/or period) their oscillation energy often gets
          re-distributed across a number of Fourier harmonics, thus
          lowering the statistical significance of QPP in the Fourier
          analysis \citep[see e.g.][for a recent review of this
            topic]{Nakariakov2019}. Hence, in this work we set two
          significance levels of 1-$\sigma$ (i.e. 68\% confidence) and
          2-$\sigma$ (i.e. 95\% confidence) in the Fourier
          spectra.\label{item:fourier}
	\item From this stage onwards, we proceed only with those
          signals which have the Fourier peaks with significance of at
          least 1-$\sigma$.
	\item For those signals, we assess the statistical
          significance of the EMD modes revealed at step
          \ref{item:emd}, following \citet{Kolotkov2016} and using the
          parameters of noise obtained from the Fourier analysis at
          step \ref{item:fourier}. The corresponding EMD spectra
          (i.e. the dependence between the EMD-revealed modal energies
          and mean periods) with the 95\% confidence levels are shown
          in the bottom right panels of
          Figs.~\ref{fig:qpp_drift}-\ref{fig:multi-modal_qpp}.\label{item:emd_confidence}
	\item We consider a QPP event is positively detected in this
          work if it has at least 1-$\sigma$ (68\%) significance in
          the Fourier analysis and 2-$\sigma$ (95\%) significance in
          the EMD analysis.
\end{enumerate}

The application of this scheme allowed us to reveal eleven QPP events
(see Figs.~\ref{fig:qpp_drift}-\ref{fig:multi-modal_qpp} and
Figs.~\ref{fig:qpp_all_1}-\ref{fig:qpp_all_3}) with mean periods
ranging from $10.2\pm1.4$\,min to $71.9\pm13.0$\,min (see
Table~\ref{looplengths}), including one QPP signal with a strong drift
in the oscillation period (see Fig.~\ref{fig:qpp_drift}) and one
two-modal QPP signal (see Fig.~\ref{fig:multi-modal_qpp}). In the QPP
event shown in Fig.~\ref{fig:qpp_drift}, the instantaneous oscillation
period is clearly seen to decrease with time from about 40~min to
20~min, unless the QPP oscillation amplitude becomes comparable to
the amplitude fluctuations caused by noise. Because of such strong
non-stationarity, the Fourier power of this oscillation is seen to be
spread between 20~min and 40~min, that resulted in overall lowering
the height of the corresponding Fourier peak below 95\% significance
level. In contrast, in the EMD method the basis for decomposition is
not prescribed a priori but is derived directly from the data by
iterative sifting its local time scales \citep{Huang1998}. This makes
the EMD method more suitable for capturing the non-stationary
oscillatory processes in general, and QPP in Solar and stellar flares
with strong period drifts in particular, see e.g. Sec.~5.4 in
\citet{Broomhall2019} and Sec.~3.3 and 4.4 in \citet{Kupriyanova2020},
and references therein.

In the example shown in Fig.~\ref{fig:qpp_drift}, the application of
EMD allowed the retaining of the energy of a non-stationary
oscillatory process seen in the detrended observational signal in a
single intrinsic mode (i.e. not distributed over a number of modes or
harmonics), that resulted in the statistical significance of this mode
above 95\% in the EMD spectrum. In the two-modal QPP event
(Fig.~\ref{fig:multi-modal_qpp}), both modes are seen to have rather
stable periods with mean values of $25.4\pm6.6$\,min and
$62.9\pm7.8$\,min, and the statistical significance about or higher
than 95\% in both the Fourier and EMD approaches. Although the
  periods of these two modes are longer than a multi-mode flare seen
  in a pre main sequence M3 star identified in NGTS data
  \citep{Jackman2019}, the ratio of the short to long periods are
  consistent to within a factor $\sim$2.
  
By making certain assumptions concerning the nature of the physical
mechanism producing the flares, various studies including those of
\citet{Mathioudakis2006} and \citet{Jackman2019} were able to estimate
the loop length of the flares where the QPPs originate. To do this, we
must first estimate the strength of the magnetic field in the M
dwarfs.

\begin{table}[!th]
\setlength{\tabcolsep}{4.9pt}
\begin{tabular}{lrrrrrcccrr}
\hline
TIC &  Energy & Period & Dur. & E-fold & $l_{\rm{slow}}$    & $l_{\rm{kink}}$  & $l_{\rm{Nam}}$   & $l/R_{*}$ & $B_{spot}$  & $B_{cor}$ \\
    &  Flare  & QPP    & & & (Mm)  & (Mm) & (Mm) & & (G) & (G)\\
    &         (erg)  & (min) & (min) & (min)    &  &      &     & & &  \\
\hline
1403938   & 6.0e35&  21.5 & 85 & 21.2 & 440 & 640 & 580 &  1.69 & 2.5k & 250\\
1403938   & 1.0e34&  36.6 & 145 & 16.1 & 750 & 1000 & 160 & 2.87 & 2.5k & 230 \\
233547261 & 7.5e35&  25.8 & 200 & 25.2 & 530 & 740 & 730 & 1.93 & 2.0k & 240\\
279494336 & 6.1e35&  39.8 & 120 & 31.3 & 810 & 950 & 760 & 2.85 & 4.0k & 200 \\
353898013 & 8.2e34&  19.8 & 78 & 20.2 & 400 & 400 & 330 & 1.76 & 1.6k & 170 \\
455825451 & 1.1e34&  71.9 & 289 & 11.9 & 1470 & 1380 & 230 &5.68 & 1.4k & 160 \\
271698144 & 3.4e35&  25.4/62.9  & 250 & 19.5 & 1810 & 1660 & 570 & 1.74 & 3.2k & 240 \\
271698144 & 4.0e34&  10.2 & 60 & 39.0 & 210 & 120 & 490 & 0.69 & 3.2k &100 \\
393804343 & 1.6e34&  13.6 & 78 & 16.6 & 280 & 230 & 280 & 0.95 & 2.0k & 140\\
393804343 & 4.3e35&  19.7 & 116 & 19.8 & 400 & 580 & 600 & 1.37 & 2.0k & 245\\
393804343 & 1.3e34&  18.5 & 140 & 32.1 & 380 & 200 & 350 & 1.29 & 2.0k & 90\\
\hline
\end{tabular}
\caption{For those stars showing QPPs we show: the flare energy; the
  QPP period; the duration of the flare based on the flux returning
  close to the pre-flare flux; the duration of the flare based on the
  e-folding time; the loop length assuming the QPP-driving mechanism
  is due to compressive standing slow magneto-acoustic oscillations,
  $l_{\rm{slow}}$; the loop length via long-wavelength kink
  oscillations, $l_{\rm{kink}}$; the loop length based on the
  predictions of the coronal magnetic field strength and loop length
  from \citet{Namekata2017}, $l_{\rm{Nam}}$ (and Fig.
  \ref{namekata}); $l_{\rm{slow}}$ compared to the star's radius;
  the estimated magnetic field of the starspot, $B_{spot}$ and the
  coronal magnetic field, $B_{cor}$. For the first QPP event in TIC
  271698144 the loop lengths are derived for the longer period QPP.}
\label{looplengths}
\end{table}

\section{Magnetic Field strengths}
\label{magfield}

Determining the
field strength (or upper limit) of any star can be achieved, in
principal, using spectropolarimetric data, but this requires
considerable telescope time and is restricted to relatively bright
stars, e.g. \citet{Reiners2012}. However, there are other indirect
means to estimate the magnetic field strength of a star provided it
shows evidence of rotational modulation. To estimate the magnetic
field strength of the stars showing QPPs we begin by using the formula
of \citet{Maehara2012} \& \citet{Notsu2019} to determine the area of
spot coverage:

\begin{equation}
\label{eq1}
\Delta {F_{\rm rot}}/F \sim \bigg[1 - \bigg(\frac{T_{\rm{spot}}}{T_{\rm{star}}}\bigg)\bigg]^{4} \frac{A_{\rm{spot}}}{A_{\rm{star}}}
\end{equation}
\vspace{1mm}

where ${\Delta} {\rm {F_{rot}/F} }$ is the amplitude of the rotational
modulation, ${\rm T_{spot}}$ and ${\rm{T_{star}}}$ are the spot and
stellar effective temperature respectively, ${\rm{A_{spot}}}$ is the
area of the star covered by the spot(s) and ${\rm{A_{star}}}$ is the
area of the star. To determine the difference between the starspot and
the mean photospheric temperature, we use the formula of
\citet{Notsu2019}:

\begin{equation}
\label{eq2}
    T_{\rm{star}} - T_{\rm{spot}} = 3.58 \times 10^{-5}T_{\rm{star}}^2 +   0.249T_{\rm{star}} - 808
\end{equation}
\vspace{1mm}

where we take the effective temperature of the star
(${T_{\rm{star}}}$) from the TIC v8.0 catalogue
\citep{Stassun2019}. The magnetic field strength can then be estimated
using the relationship derived by \citet{Shibata2013}:

\begin{equation}
\label{eq3}
  E_{\rm{flaremax}} = 7\times10^{32} \rm{erg} \left( \frac{\it f}{0.1}  \right)  {\left( \frac{\it B}{10^3} \right)}^2 {\left( \frac{A_{\rm{spot}}/2\pi {\it R}^2}{0.001} \right)}^{3/2} 
\end{equation}
\vspace{1mm}

where $E_{\rm{flaremax}}$ is the maximum bolometric energy of the
flare we detect in that stars light-curve, $f$ is the fraction of the
magnetic energy that can be released as flare energy (which we fix at
0.1 as done by \citet{Shibata2013}) and $A_{\rm{spot}}$ is the
area of the spot (taken from Equation \ref{eq1} above).

\begin{figure}
  \begin{center}
  \includegraphics[width=0.87\textwidth]{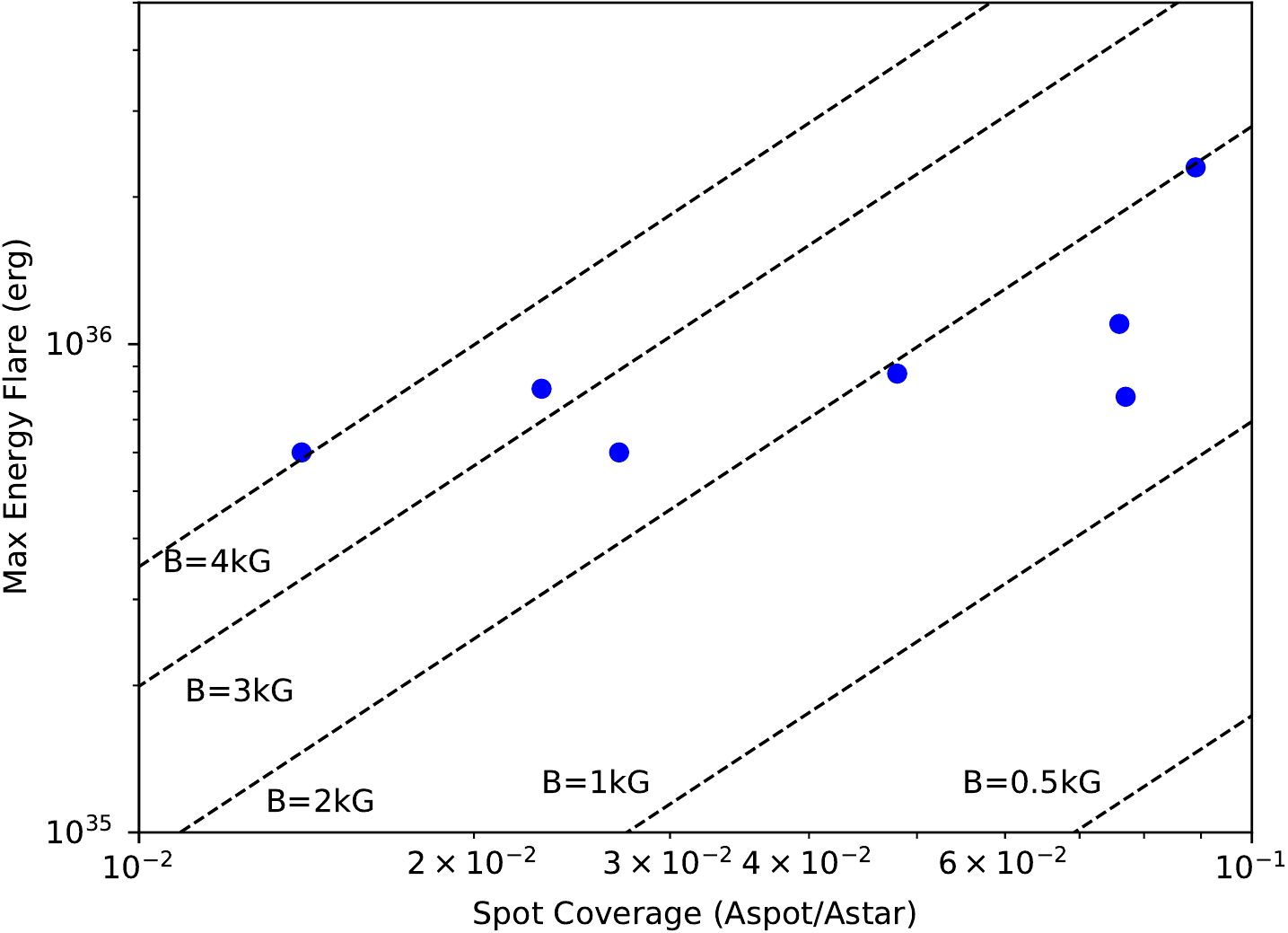}    
\vspace{2mm}
  \caption{The relationship of spot coverage and the bolometric energy
    of the most energetic flare from that star. We show the predicted
    relationship for different magnetic field strengths from equation
    \ref{eq3} and \citet{Notsu2019}.}
    \label{spotcoverage}
    \end{center}
\end{figure}

Using the equations above we can determine the relationship between
the relative size of the starspot and the maximum energy of the flare
which we show in Fig. \ref{spotcoverage}. Using equation \ref{eq3} we
can derive the magnetic field of the starspot as a function of area
and energy and find $B\sim$1--4~kG.  These field strengths are consistent
with the magnetic field of starspots in M dwarf stars, e.g. see
\citet{Morin2008}.

\section{Loop lengths}
\label{looplength}

\citet{Roberts1984} outlined how the QPPs seen in Solar flares could be
used to determine the loop length of Solar coronal loops and the
physical conditions in their immediate environment. The same
principles have been used to determine properties of stellar flares
using observations of QPPs. \citet{Mathioudakis2006} approximate the
formula for the period and loop length to:

\begin{equation}
\label{eq4}
  {\rm Period (sec)} = \frac{l_{\rm{slow}}~{\rm(Mm)}}{7.6\times10^{-2} N \sqrt{T \rm{(MK)}}}
\end{equation}
\vspace{1mm}

where $l_{\rm{slow}}$ is the loop length in Mm, $N$ is the node of
oscillation ($N$=1 for the fundamental, $N$=2 for the first harmonic)
and $T$ is the average temperature of the corona along the flare loop
which \citet{Mathioudakis2006} take to be 20~MK. The above equation
assumes that the QPP-driving mechanism is due to compressive standing
slow magnetoacoustic oscillations in coronal loops where the
oscillation period is prescribed by the loop length and the sound
speed (i.e. square root of temperature).

Assuming $N$=1, we show the derived loop lengths in Table
\ref{looplengths} which are in the range $\sim$200--1800~Mm (they are
twice these values if $N$=2). Taking the radius of each star from the
TIC \citep{Stassun2019}, we find that the loop lengths are typically
of the same extent (or greater) as the stellar radius. This is
consistent with a study of 44 stars with F-M spectral type which
showed that the loops with the largest length to stellar radius ratio
($\sim2 R_{*}$) originated on M dwarfs \citep{Mullan2006}, with the
shortest loops being comparable to that estimated for the flare
reported in \citet{Jackman2019}.

\begin{figure}
  \begin{center}
  \includegraphics[width=0.87\textwidth]{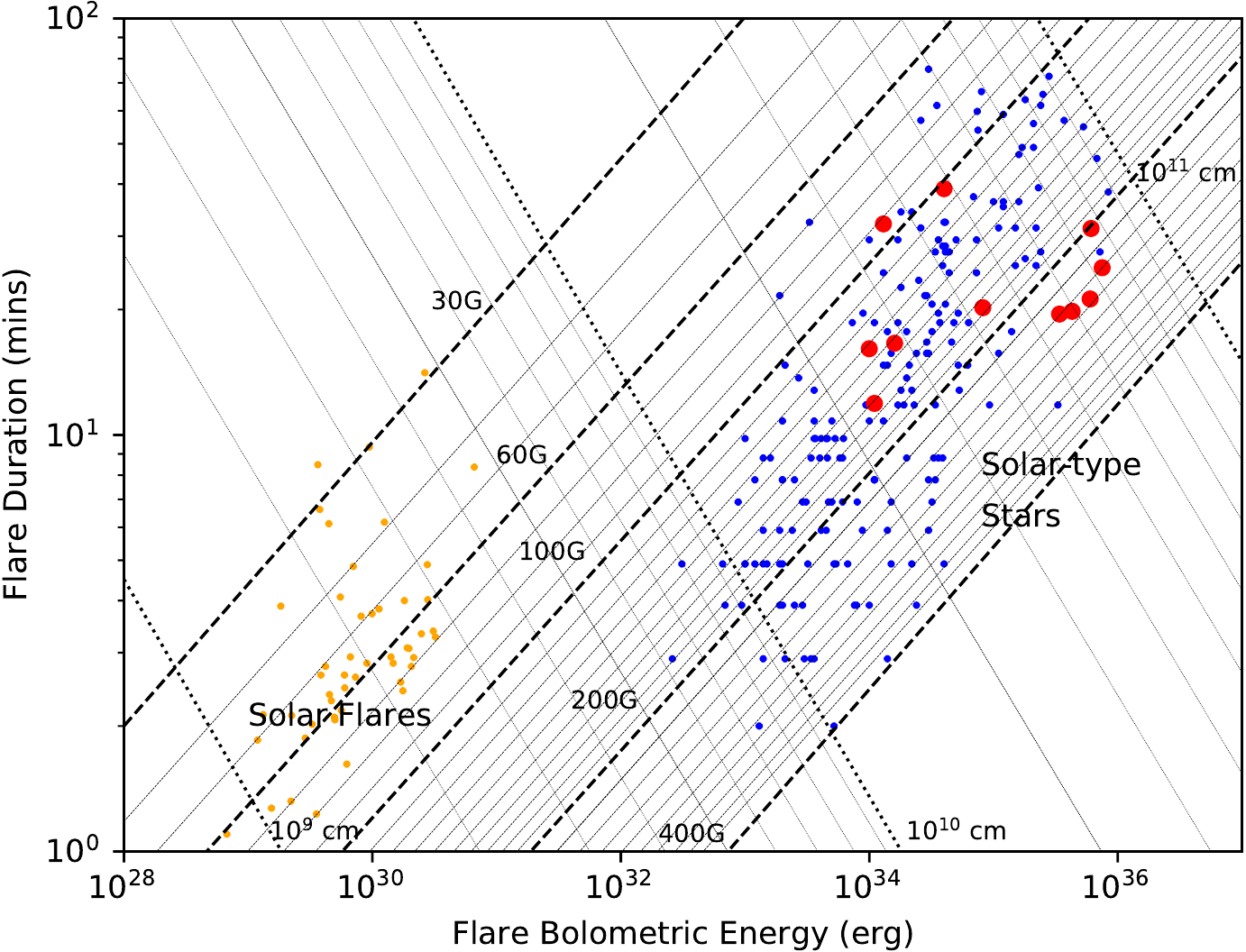}    
\vspace{2mm}
  \caption{This figure is based on Fig 9 of \citet{Namekata2017} which
    incorporate work from \citet{Maehara2015}.  Solar data is shown as
    orange dots whilst blue dots show super-flares from Solar type
    stars using {\kep} data with 1 min cadence. The large red dots are
    the sources shown in this paper where the energies and duration
    are derived from {\tess} data. We can use the relationships
    derived for different magnetic field strength and loop length to
    estimate these quantities for the stars presented in this paper.}
    \label{namekata}
    \end{center}
\end{figure}

An alternative way of deriving the loop length is via the scaling laws
used by \citet{Namekata2017}. Taking the flare duration and energy of
those flares as outlined in Table \ref{looplengths}, we place them in
context by adding them to Fig. 9 of \citet{Namekata2017} which shows
the energy and duration of Solar flares and Solar-like stars. Our
sample of low mass stars are comparable with the more energetic flares
from Solar-like stars. We can estimate the loop lengths of the flares
in our sample by comparing their location in Fig. \ref{namekata}
with the theoretical relationships for coronal magnetic strength and
loop length which were taken from \citet{Namekata2017}. Given the
uncertainties in the assumptions, the resulting loop lengths are
similar to those estimated assuming they are driven by slow
magneto-acoustic oscillations.  However, no matter which wave mode we
use, these loop lengths estimates clearly imply very large active
regions covering a significant factor of the stellar surface.

However, alternative QPP-driving options are possible, e.g. fast
magnetoacoustic waves which include the {\it kink oscillations} of
coronal loops situated in, or nearby, the flaring active region
\citep{Nakariakov2006}. The period of standing kink oscillations is
prescribed by the loop length and the Alfv\'en speed inside the loop
\citep{Nakariakov2020}. Here we have two unknowns, the loop length and
the Alfv\'en speed. In \S \ref{magfield} we estimated a photospheric
magnetic field typically in the range $\sim$1--4~kG. There are obvious
uncertainties in predicting the coronal magnetic field strength from a
photospheric field strength (the uncertainty in the height of the loop
above the photosphere and the magnetic field configuration) but a
reduction of a factor of 10 is reasonable. This yields coronal
magnetic field strengths roughly between $\sim$100-300~G (see Table
\ref{looplengths}) which are in excellent agreement with those derived
from Fig.  \ref{namekata}. We can use the equation

\begin{equation}
\label{eq5}
{\rm Period} = \surd{2}l_{\rm{kink}}/C_{\rm{Alfv\acute{e}n}}
\end{equation}
\vspace{1mm}

where $l_{\rm{kink}}$ is the loop length and $C_{\rm{Alfv\acute{e}n}}$
the Alfv\'en speed based on the derived magnetic field strength,
assuming the fundamental (global) harmonic (see Eq. 9 of
\citet{Nakariakov2020}) and a density $\rho_{o}=1\times10^{-9}$
  kg m$^{-3}$ which implies an electron number density, $N_{e}\sim1\times10^{12}$
  cm$^{-3}$ (\citet{MonsignoriFossi1996} found that in the corona of
  active M dwarfs, $N_{e}>1\times10^{12}$ cm$^{-3}$ during activity
  and $N_{e}\sim1.5\times10^{13}$ cm$^{-3}$ during flares).  The
  derived loop lengths (Table \ref{looplengths}) are in good
  agreement with the lengths derived via the compressive standing slow
  mode. The sound speed $C_{\rm s}$ is about the Alfv\'{e}n speed
$C_{\rm Alfv\acute{e}n}$, i.e. the plasma parameter $\beta=2C_{\rm
  s}^2/(\gamma C_{\rm{Alfv\acute{e}n}}^2)$ is around unity. In Solar
flares, such high values of $\beta$ have been observed, e.g. high
energy X-class flares require high temperatures and that both the
plasma $\beta$ and volume filling factor cannot be much less than
unity in the super-hot region, see \citet{Caspi2014}. 

We now comment briefly on the parameters for flare energy, duration,
loop lengths and magnetic field reported in Table
\ref{looplength}. Given the loop lengths are directly linked to the
period of the QPP (Eq. \ref{eq4}) there is a clear correlation between
these parameters and also period and duration: long duration flares
can have long period QPPs whilst short duration flares
cannot. Similarly, the correlation between magnetic field strength and
duration of the flare comes from the scaling relations of
\citet{Namekata2017}. Perhaps surprisingly we find no correlation
between the flare energy and flare duration.

\section{Flare rates and the effects on potential Exoplanets}
\label{flarerates}

\begin{figure*}
  \begin{center}
  \includegraphics[width=0.89\textwidth]{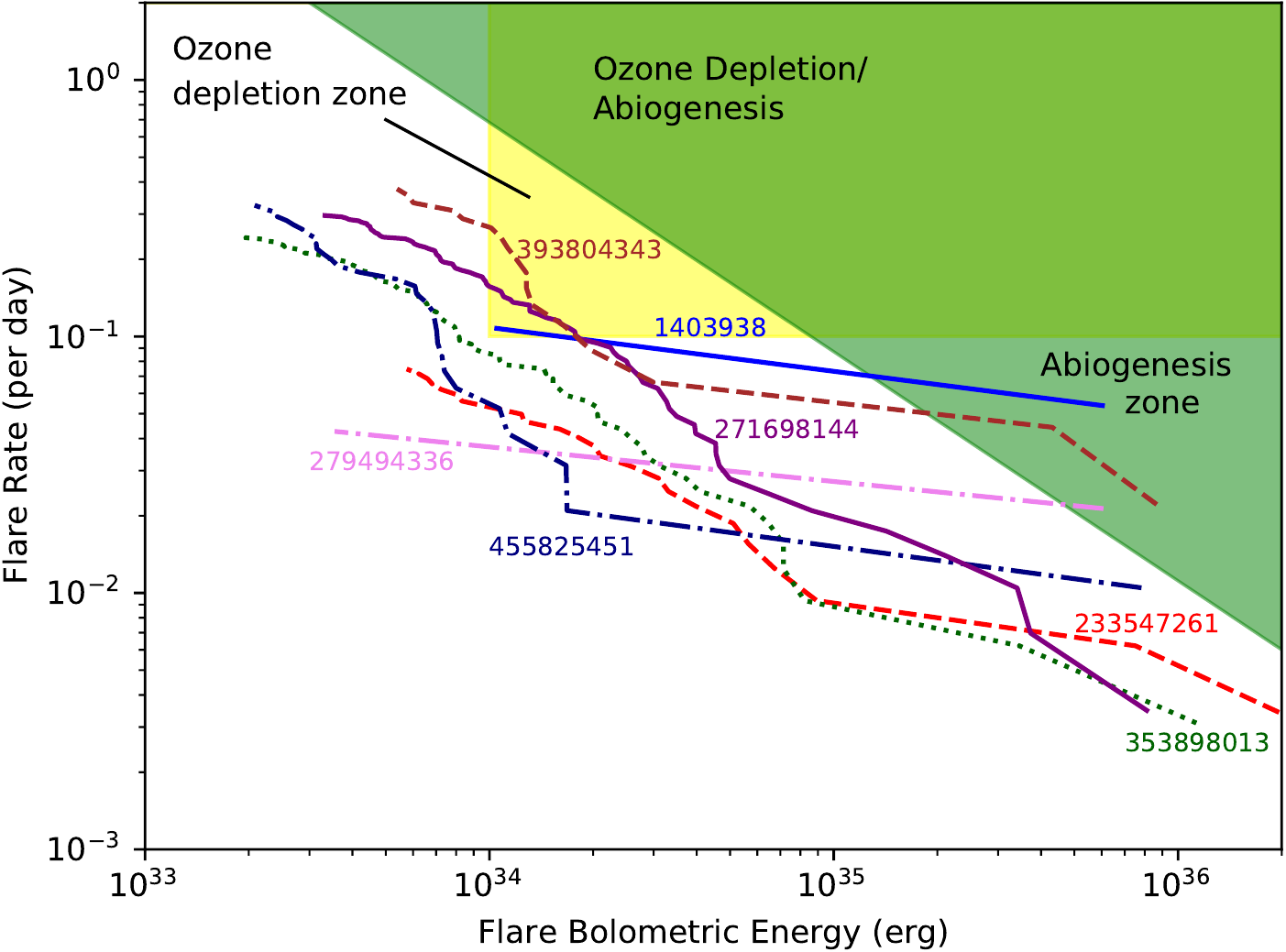}    
\vspace{2mm}
  \caption{The Flare Frequency Distribution (FFD) for flares of given
    energy for all seven stars in our sample which show QPPs. The
    Solar FFD, during both maximum and minimum, have maximum energies
    which are lower than the minimum energy shown here. We show
      the regions where terrestrial exoplanets in the habitable zone
      will have ozone in their atmosphere depleted and also lie in the
      abiogenesis zone, which were taken from Figure 11 of
      \citet{Gunther2020} which was adapted from work of
      \citet{Tilley2019} and \citet{Rimmer2018}.}
    \label{energyrate}
    \end{center}
\end{figure*}

We now return to the overall flare rates of the seven stars for which
we have identified QPPs. We show these rates in Fig. \ref{energyrate}
as a cumulative flare frequency distribution (FFD). The seven stars
show flares with bolometric energies $>10^{35}$~erg occurring at a
rate of 1 per $\sim$10--100 days: these are higher rates than the
average rates for M3V-M4V stars \citep{Howard2019}.

We do not examine the physical mechanism which enable these stars to
produce high energy flares at such high rates. Rather, we examine what
effect these flares may have on the atmosphere of any orbiting
exoplanet. Over the past 25 years nearly 4,300 exoplanets have been
discovered with thousands more awaiting confirmation. As a result, we
now know the majority of main sequence stars are likely to host
planetary systems. Furthermore, the study of stellar activity
(including flares) on these host stars has become particularly
prominent in exoplanet research. This is due to the impact on the
exoplanet atmosphere and the existence for life along with the
potential of exoplanet signals being masked by stellar variability.

It is widely known that for life to exist, the planet must lie within
the habitable zone (HZ) of the star. This is a planets distance from the
star where liquid water is likely to pool on the surface. However,
this is not the only requirement for life. In \citet{Rimmer2018}, they
use experimental chemistry along with stellar physics to determine how
much energy would be needed for abiogenesis (natural process where
life arises from non-living matter) to occur. They compute this for a
sample of low mass stars from \citet{Davenport2016}, concluding that
$\sim$20\% of early M dwarfs are active enough for any host planets to
be in the abiogenesis zone. The abiogenesis zone indicates where the
stellar UV flux is large enough to result in a 50\% increase of the
photo-chemical product (i.e. those products which are present in the
nucleotide synthesis pathway, see Fig. 1 of \citet{Rimmer2018} for more
details).

Furthermore, there is an ozone depletion region where flare rates can
cause ozone loss for a planet orbiting in the HZ around an
M dwarf. Understanding the effects of flaring activity on an exoplanet
atmosphere is essential in predicting whether there is life on the
surface of the planet. The exoplanet could well be in the HZ, however,
it may also be bombarded by electromagnetic and particle radiation
from the star. \citet{Tilley2019} looked into this in more detail
using models to investigate the effects of repeated flaring on the
photochemistry and surface UV of an Earth-like planet (unprotected by
a magnetic field) orbiting an M dwarf. Overall, they found 
Coronal Mass Ejections (CMEs), which are commonly associated with flares
possessing energies $>10^{34}$~erg, were the primary factor in ozone
depletion, at a rate greater than once every ten
days. The ozone is particularly important for the existence of
surface life as this thin layer is responsible for absorbing almost
all of the harmful UV radiation. 

We have taken the location of the Ozone depletion and Abiogenesis
  zones from Fig. 11 of \citet{Gunther2020}, who have determined the
  abiogenesis zones for FGKM stars and for flare energies in the $U$
  and the bolometric energies (as we have derived). Specifically we
  use the regions determined for M0-M4 stars.  As can be seen from
Fig. \ref{energyrate}, three stars have FFDs which overlap with
  the ozone depletion zone. This means any potential exoplanets
  orbiting these stars could have their ozone layer destroyed by the
  constant flaring activity. The frequency of this high energy
flaring means the atmosphere of an exoplanet is unlikely to have time
to recover, being destroyed over a prolonged period of time assuming
the absence of a magnetic field environment. Although, whether a
magnetic field adds sufficient protection remains unanswered. For
example, magnetised Earth has a relatively thick atmosphere whereas
Mars, which is unmagnetised, does not. However, on the other hand,
Venus which is also unmagnetised has a relatively thick and dense
atmosphere. Regardless of this, the UV radiation exposed to any
potential exoplanets would make it difficult for surface life to
exist. This does not rule out the existence of life altogether as
there could be some present below the ocean surfaces.

There are three stars in the sample which have FFDs which extend
  to the abiogenesis zone: TIC 393804343, TIC 1403938 and TIC
  279494336. For these stars this is significant as it indicates the
  high energy flares of $\sim5\times10^{35}$~erg at a frequency of one
  every $\sim$20--50~days, is an important factor in the potential for
  life on any orbiting exoplanets. However, there is a fine line
where too many high energy flares at a high frequency can cause ozone
depletion. We see an example of this in TIC 1403928 and TIC 393804343
which lie in both the ozone and abogenesis zones. The question for
this particular star is whether any orbiting planets atmosphere will
have enough time to recover, allowing for abiogenesis and surface life
to potentially exist. The remaining stars do not lie in either
  the abiogenesis or ozone depletion zones, although they do show high
  energy flares but at rates slightly lower than required to be in the
  abiogenesis zone. As a result, the ozone layer on any orbiting
planets will likely remain intact and the UV flux is not high enough
to sustain prebiotic chemistry. This does not mean there is no
potential for any surface life on orbiting planets, just that it is
less likely to manifest without the natural environment needed.

Most M dwarfs have not been observed in EUV/UV. However,
\citet{Melbourne2020} derived a relationship between Ca II H\&K and
H$\alpha$ with various strong FUV/NUV lines. The derived scaling
relations may be sufficient for photochemical modelling needs, but not
atmospheric escape modelling, although an accurate estimate of the
Ly~$\alpha$ flux should account for $\sim$75\% of the FUV in the
quiescent state and $\sim$50\% during a flare
\citep{Diamond-Lowe2021}. Unfortunately, we do not yet have the Ca II
H\&K or H$\alpha$ flux for these objects.

In Fig. \ref{qpp_schematic}, we show a schematic plot of how far the
coronal loops extend out from the stellar surface in relation to the
HZ. In determining this, we assume the loops are semi-circular arcs
which are equal to the loop length. As a result, the radius of the arc
is then the distance at which the loop extends out from the stellar
surface and can be calculated as $R_{\rm{loop}} = l/\pi$. It is
important to note that this yields a conservative estimate on the
radius of the loop, as most loops are not perfect semi-circular arcs
but appear more elliptical in shape. However, for the purposes of this
paper assuming a semi-circular arc is acceptable. For stars which
possessed more than one QPP event in Table \ref{looplengths} the QPP
with the largest loop length was used. The HZ boundaries were
calculated using \citet{Kopparapu2013,Kopparapu2014} taking the
effective stellar temperature, $T_{\rm{eff}}$, and luminosity, $L$,
from Table \ref{stellar_properties}.

It is important to note that none of the targets in this sample have
any known exoplanets according to the NASA Exoplanet
Archive\footnote{\url{https://exoplanetarchive.ipac.caltech.edu/}}. This
does not mean there are none present at all, just that none have been
detected around these stars as of yet. In fact, the occurrence rate
for exoplanets orbiting mid-M dwarfs is on average 1.2 planets per
star \citep{Hardegree-Ullman2019} so, each of these targets is
expected to have at least one orbiting exoplanet.

\section{Discussion}

The promise of finding QPPs in {\tess} observations of low mass dwarf
stars was demonstrated in observations of Proxima Centauri which
showed oscillations on a period of several hours in the decay phase of
two flares \citep{Vida2019}. The work presented here shows the scope
for identifying and analysing QPPs from a larger sample of low mass
stars using {\tess} data. 

Of the 178 M dwarfs which we detected at least one flare with an
amplitude corresponding to $>$0.5 mag and duration longer than 1 hr,
we found 11 flares from 7 stars which showed QPPs.  Therefore only 4\%
of the stars showed flares with QPPs. 

However, of those stars which {\sl did} show QPPs, we find a very
diverse rate for the fraction of flares which showed QPPs. For
instance, QPPs were detected in both of the two flares seen in TIC
1403938. In contrast, for TIC 353898013 only 1 out of 79 flares showed
a QPP. However, those stars with relatively high fractional rates also
has the shortest duration of observations so this may simply be an
observational bias.

In several instances, the flare showing QPPs was not the most
energetic event of that star, which raises a question of the
excitation of waves in stellar atmospheres by powerful flares, to be
addressed in follow up work.  However, we note that 11 QPP events
detected in this work constitute 7\% of the total number of flares
examined (150), which is a factor of two greater than the statistics
of stellar QPPs previously reported \citep{Balona2015,Pugh2016}, and
coincide with the fraction of Solar C-class flares which show QPPs
\citep{Hayes2020}.
We now briefly discuss the robustness of the loop lengths we derive,
the effect of QPPs on any orbiting exoplanets and look ahead to future
observations.

\subsection{Loop Lengths and Scaling Laws}

\begin{figure*}
    \centering
    \includegraphics[width = 1.00\textwidth]{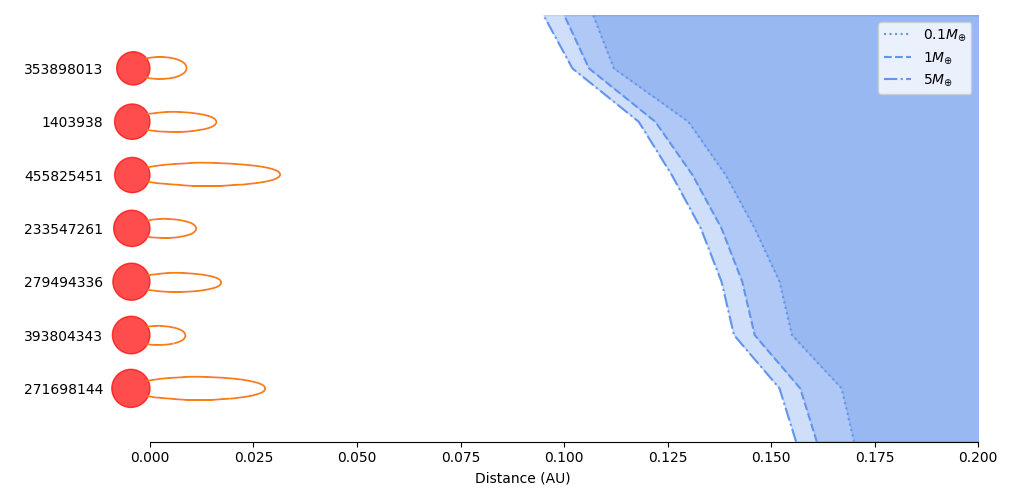}
    \caption{A schematic diagram showing how far the coronal loops
      (compressive standing slow mode) extend out from the stellar
      surface, relative to where the HZ lies from the star. The inner
      HZ boundary is shown for 0.1 (dotted), 1 (dashed) and
      5$M_{\oplus}$ (dashed-dot) and the outer boundary lies between
      0.2 and 0.3~AU (not shown in the diagram). These HZ boundaries
      were calculated using \citet{Kopparapu2013, Kopparapu2014} and
      loop lengths ($l_{\rm{slow}}$) were taken from Table
      \ref{looplengths}.}
    \label{qpp_schematic}
\end{figure*}

As outlined in \S 1, there are over a dozen possible QPP
mechanisms. However since, we are dealing with oscillations with
periods of tens of minutes, most of these are not possible. In each of
the QPP mechanisms discussed in \S \ref{looplength} we have made a number
of assumptions. For example, with the slow mode waves we assume a
coronal flare temperature of 20~MK; using say 10~MK instead will
decrease the estimate by $\sim$30\% ($\propto \sqrt{T})$. For the
kink waves, there is some uncertainty regarding the plasma density. 
In flare conditions,
the density can be high; e.g. an order of magnitude
increase will decrease the Alfv\'en velocity and the loop length by a
factor of three. Considering the various approximations and
assumptions used in all three loop length calculations, the derived
lengths can only be considered good to an accuracy of a factor of
two. Nevertheless, the derived lengths are all comparable with the
star's radius and as shown in the illustration in Fig. 
\ref{qpp_schematic}, these loops extend high into the star's corona,
thus providing intense UV radiation into a planet's HZ.

Whilst we considered the kink and slow magnetoacoustic modes of a
  coronal loop as the most common and straight forward mechanisms for
  QPPs, a more detailed discussion of this question is beyond the
  scope of this work. Indeed, the problem of the association of the
  observed QPP signals to the wave and oscillatory phenomena in solar
  and stellar atmospheres has been actively debated for at least the
  last fifteen years and still remains an open question
  \citep{Zimovets2021}. To address the question of which mechanism
  drives the QPP events and to gain a better understanding of the
  physical mechanism producing white light flares will require many
  more QPPs to be detected and studied.

\subsection{Possible Effects of QPPs on Orbiting Exoplanets?}

Since the launch of {\tess}, the discovery of planets orbiting M dwarf
host stars has rapidly increased. These low mass stars spend billions
of years on the main sequence due to their cooler temperature. As a
result, this would provide enough time for life to flourish on the
surface of orbiting exoplanets making the study of exoplanet
habitability of particular importance. In section \ref{flarerates}, we
discussed the flare energy and rate of our M dwarf sample and whether
any of the stars possess the conditions needed for abiogenesis or
ozone depletion. However, we did not discuss the effects of QPPs on
the atmosphere of an exoplanet. \citet{Hayes2017} investigated the
effects of Solar flare X-ray QPPs on the Earth's atmosphere. They
found pulsations within the lower ionosphere of the Earth matched
those of observed Solar QPPs. This QPP-driven periodic modulation of
the conditions in Earth's lower ionospheric (e.g. the electron number
density) could result in phenomena such as acoustic gravity waves
  (e.g. \citet{Nina2013}), and disruptions in high-frequency radio
  wave propagation (e.g. \citet{Frissell2019}), implying a similar
  response could be found in exoplanet atmospheres.

If a planet was orbiting close enough to the host star then
magnetospheric interactions could be possible. This is when the host
star and close-in planets magnetic fields interact causing Star Planet
Interactions (SPIs) in the form of increased chromospheric emission
and possibly flaring (see, \citet{Ip2004,Cohen2010,Strugarek2014}. In
our stars, we have coronal loops which extend out to between 0.008 to
0.03~AU (see, Fig. \ref{qpp_schematic}). None of the stars in our
sample have known exoplanets, however, there have been exoplanets
detected around other low mass stars within this range \citep[for
example the TRAPPIST-1 system][]{Gillon2017}. TRAPPIST-1 has seven
planets orbiting an M8 dwarf star with semi major axis between 0.01
and 0.06~AU. This means that at least four of the planets are orbiting
the star within the range of the coronal loops we observe on our
sample. Therefore, it is possible to have planets orbiting within the
loops, outwith the HZ, causing SPIs with the host star. The effects
this would have on the planet would make it unlikely for any surface
life to exist due to an increase in flare activity resulting in an
onslaught of harmful radiation.

\section{Future Observations \& Conclusions}

The {\tess} satellite completed its primary mission (covering Cycle 1
and 2) in July 2020 during which it covered the majority of the
sky. In each hemisphere, there was an area at each ecliptic pole known
as the continuous viewing zone where stars could be observed for a
year (some gaps are present due to roughly monthly field
re-pointing). During Cycle 1 and 2, around 20,000 stars were observed
in 2 min cadence mode with 30 min cadence data being available for all
stars through full-frame image data.

The extended mission started in July 2020 by re-observing the southern
ecliptic hemisphere. There was, however, a clear difference in the
cadence of targets. Around 600 targets will be made available to be
observed in 20 sec cadence through the Guest Observer Programme and
15,000 targets will be observable in 2 min cadence mode. The
full-frame images will now be made available with a cadence of 10 min.

Bright low mass dwarfs which have already been shown to be flare
active, will be excellent targets to observe in 20 sec cadence mode as
it will provide increased time resolution of QPP events. This is
especially true for shorter period QPPs which will possess shorter
loop lengths. For long duration flares, the full frame data will prove
useful in extending the sample of stars showing QPP events.

Stars which have shown QPPs in {\tess} data will be excellent targets
to be observed using high cadence ground based imagers such as
ULTRACAM. Previous observations made using ULTRACAM have allowed the
colour, and hence temperature, of the flare to be determined over the
event, which have provided the base material to test competing models
which give rise to the QPP events \citep[e.g.][]{Kowalski2016}.  The
derived loop sizes from the QPPs is consistent with work by
\citet{Cohen2017} who found that at high latitudes the corona and its
X-ray emission are dominated by star-size large hot loops. These
authors suggested that in rapidly rotating stars, emission from such
coronal structures dominates the quiescent saturated X-ray emission.

The work presented here shows the scope for identifying and analysing
QPPs from a larger sample of low mass stars. We found 7 stars which
show a total of 11 QPP events, one of which is a double mode
event. With future TESS 20 sec cadence data, we can expand on this
work sampling shorter periods which will allow a better comparison
with Solar data.  For example, \citet{Hayes2020} analysed 5,519 Solar
flares observed in the X-ray, covering X-, M- and C-class events in
the past Solar cycle. Looking at periodicities in the 6-300~s
timescale the authors found QPPs in 46\% of X-class, 29\% of M-class,
and 7\% of C-class flares. The data used here does not allow us to
search this period range, but future TESS data will allow us to look
for periods down to $\sim$ 60~s and in particular look at events with
energy classifications ranging up to X100,000; an energy range not
possible for Solar flares.

Using the observed properties of the stars we estimate the length of
the loops giving rise to the flares: they are typically of the order
of the stellar radius. The observed properties of QPPs are found to be
consistent with the interpretation in terms of magnetoacoustic waves
potentially present in the atmospheres of the observed stars. However,
we could not discriminate between the fast or slow magnetoacoustic
modes because they have rather similar observational properties in
high-$\beta$ flaring plasmas. We fully expect other QPPs to be
identified in future {\tess} data.

\section*{Acknowledgements}

This paper includes data collected by the {\tess} mission. Funding for
the {\tess} mission is provided by the NASA Explorer Program.  This
work presents results from the European Space Agency (ESA) space
mission {\sl Gaia}. {\sl Gaia} data is being processed by the {\sl
  Gaia} Data Processing and Analysis Consortium (DPAC). Funding for
the DPAC is provided by national institutions, in particular the
institutions participating in the {\sl Gaia} MultiLateral Agreement
(MLA). The Gaia mission website is
\url{https://www.cosmos.esa.int/gaia}. The Gaia archive website is
\url{https://archives.esac.esa.int/gaia}. Armagh Observatory and
Planetarium is core funded by the N. Ireland Executive through the
Dept. for Communities. D.Y.K. was supported by the STFC consolidated
grant ST/T000252/1 and by the Ministry of Science and Higher Education
of the Russian Federation. L.D would like to acknowledge funding from
a UKRI Future Leader Fellowship, grant number MR/S035214/1. We thank
Richard West for raising the possible issue of chromatic aberration
and the anonymous referee for a useful report.

\vspace{4mm}

\bibliographystyle{spr-mp-sola}

\appendix
\section{QPP signals}
\begin{figure*}
	\centering
	\includegraphics[width=\linewidth]{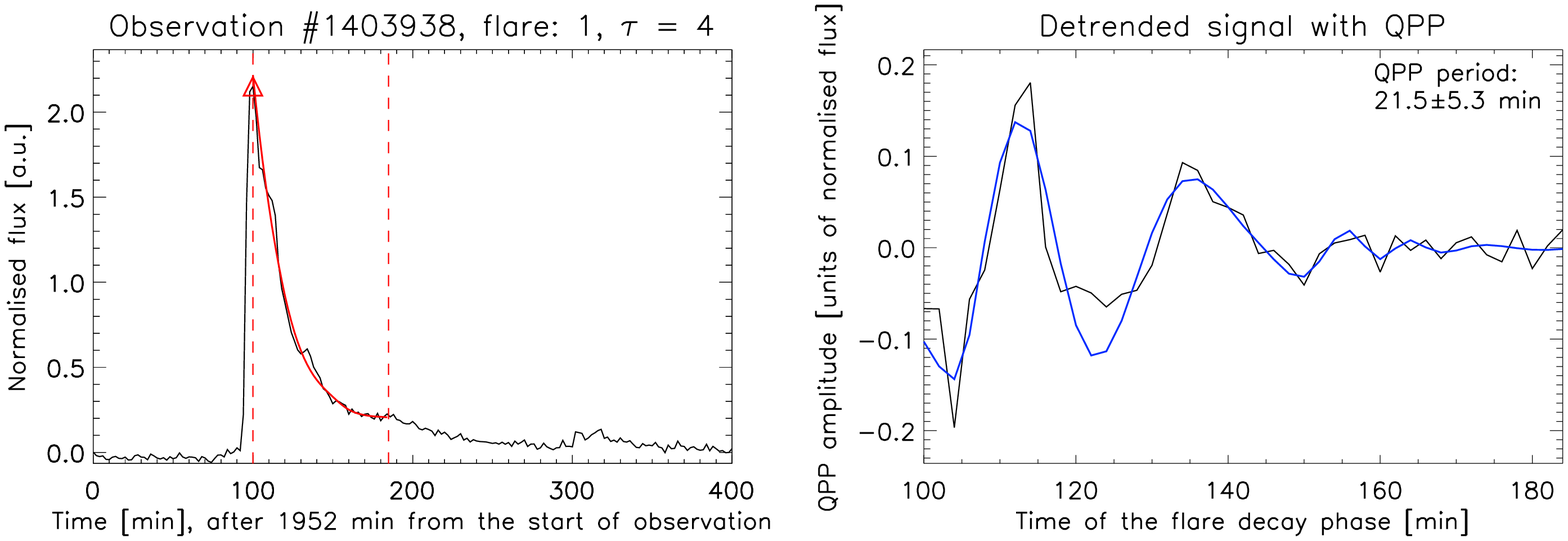}
	\includegraphics[width=\linewidth]{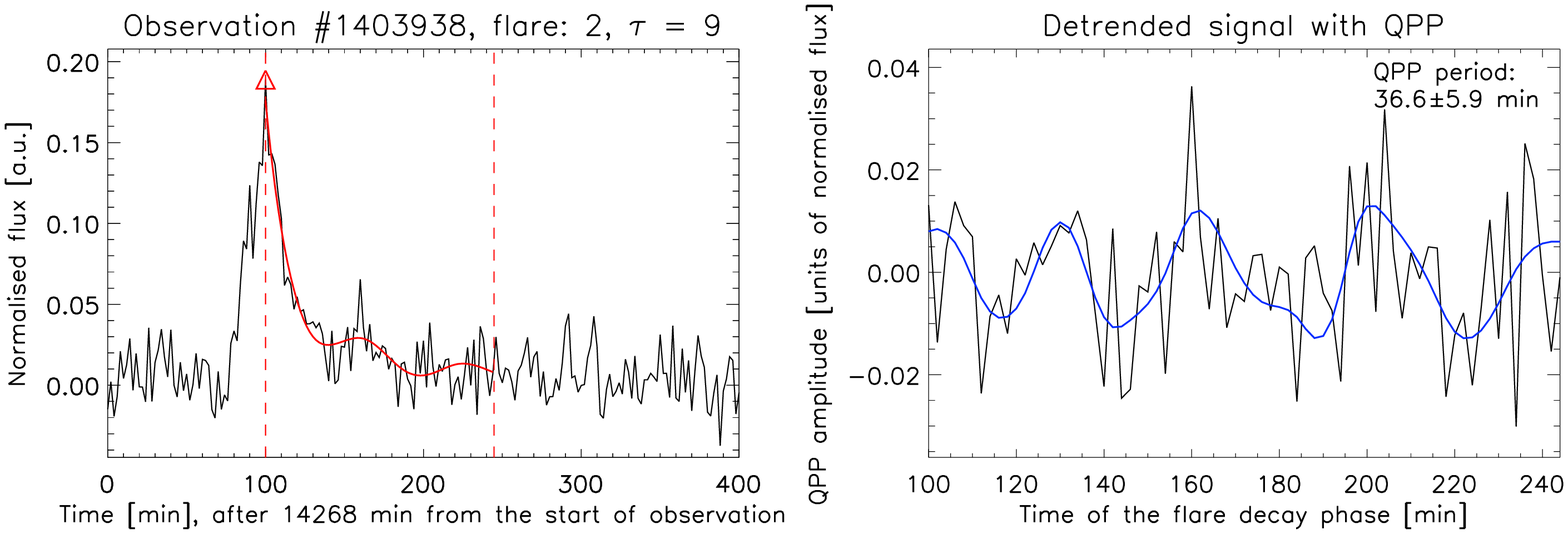}
	\includegraphics[width=\linewidth]{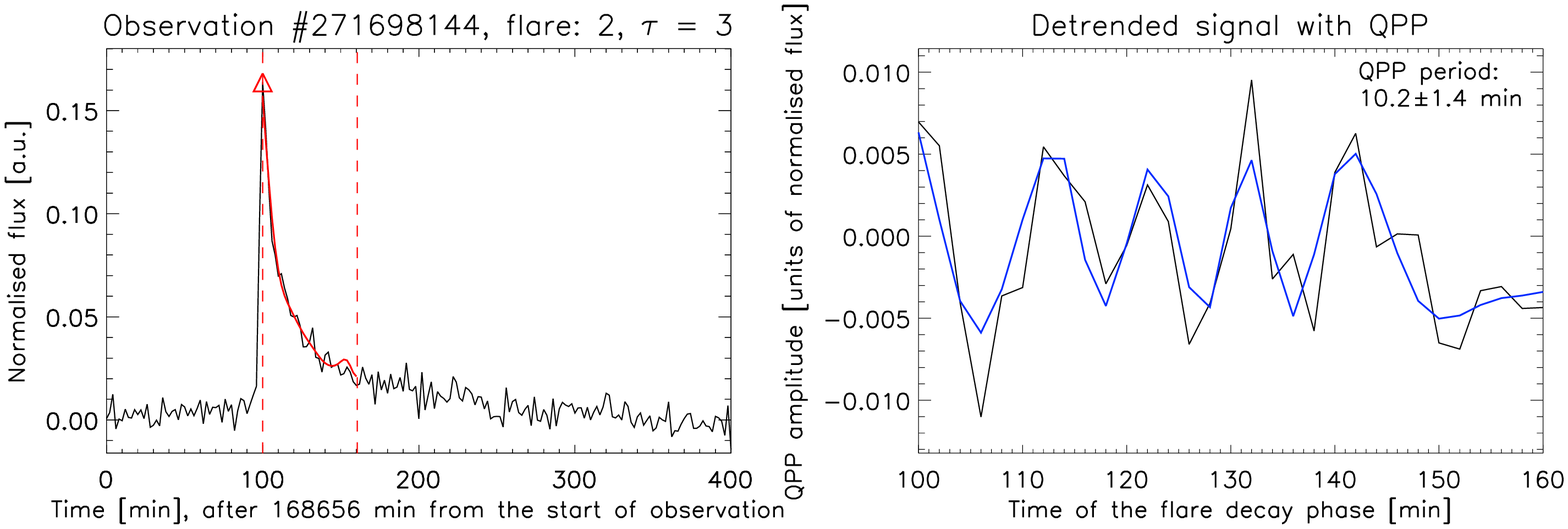}
	\caption{The QPP signals detected in observations TIC 1403938 and TIC 271698144. See Fig.~\ref{fig:qpp_drift} for details on the figure notations.
	}
	\label{fig:qpp_all_1}
\end{figure*}

\begin{figure*}
	\centering
	\includegraphics[width=\linewidth]{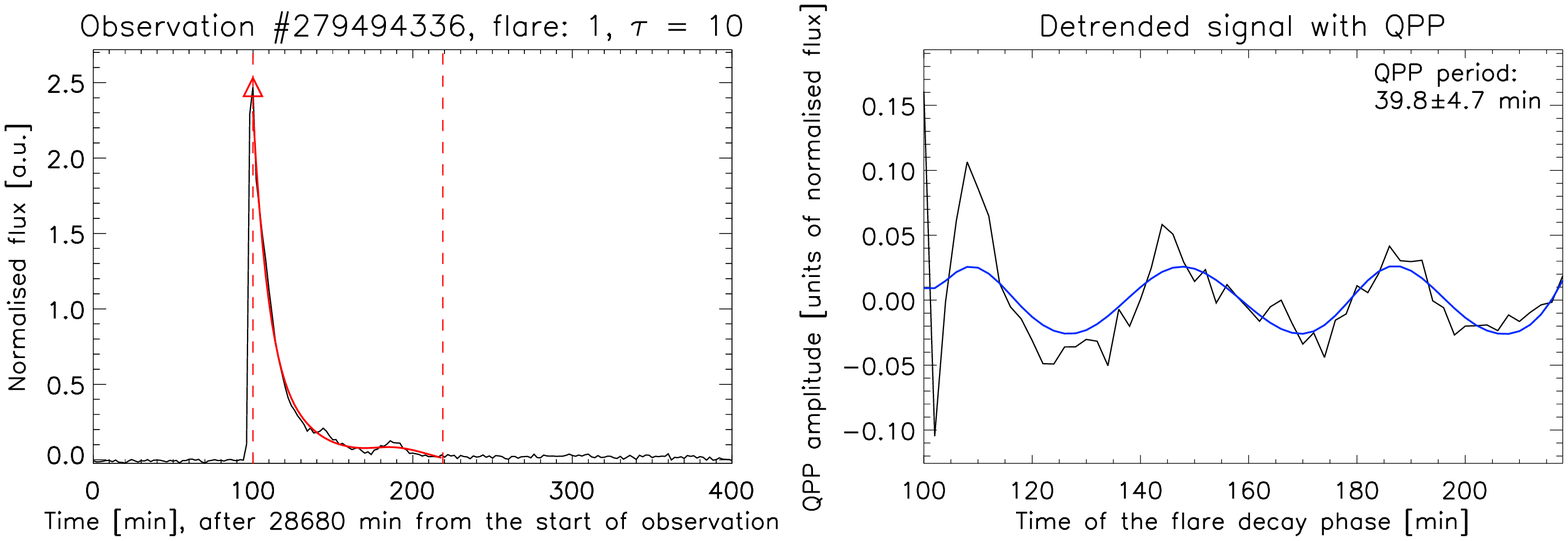}
	\includegraphics[width=\linewidth]{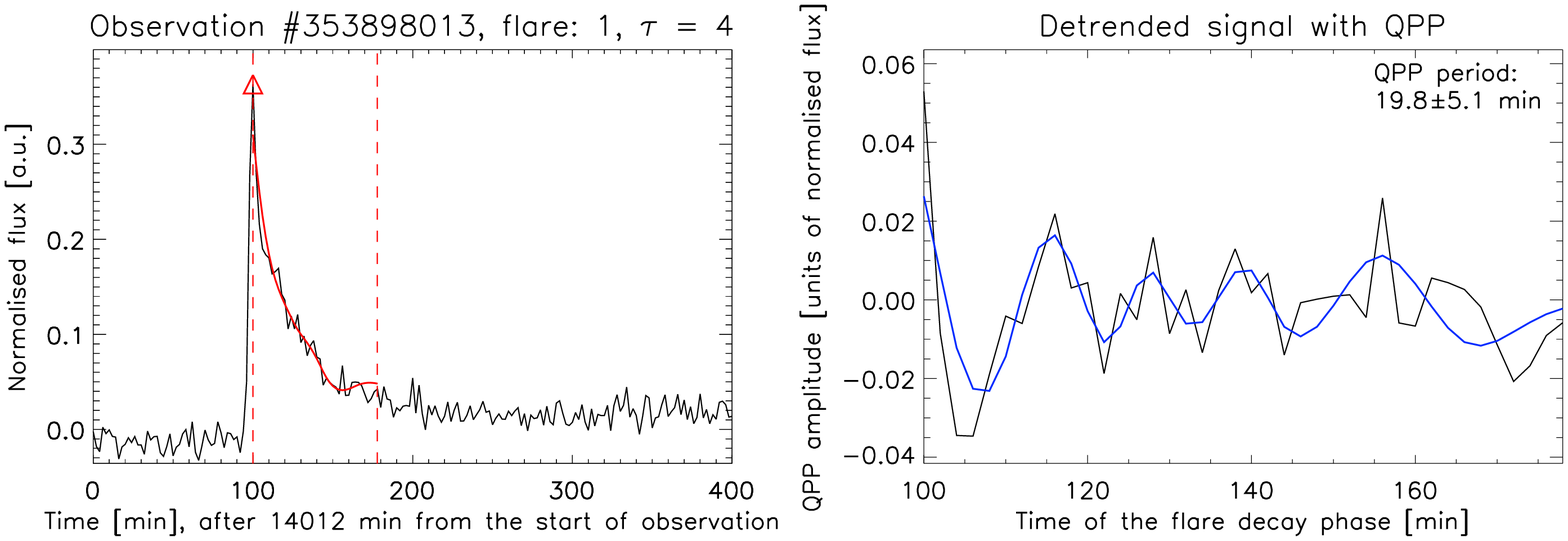}
	\includegraphics[width=\linewidth]{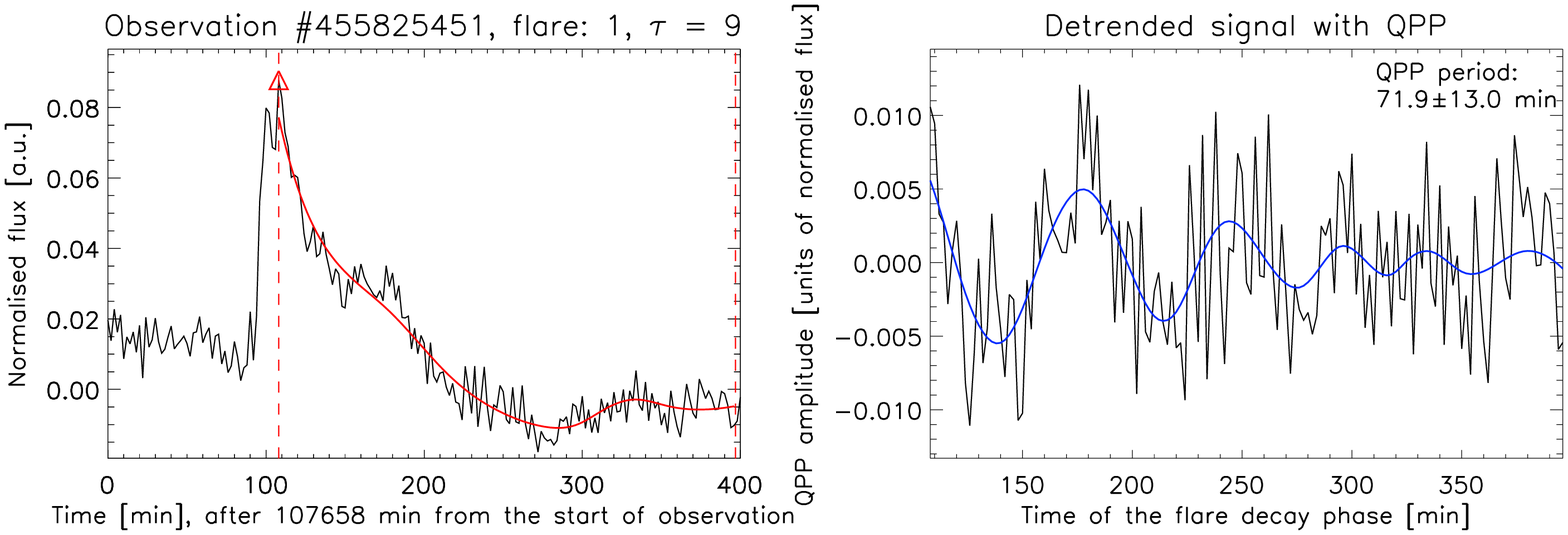}
	\caption{The QPP signals detected in observations TIC 279494336, TIC 353898013, and \#455825451. See Fig.~\ref{fig:qpp_drift} for details on the figure notations.
	}
	\label{fig:qpp_all_2}
\end{figure*}

\begin{figure*}
	\centering
	\includegraphics[width=\linewidth]{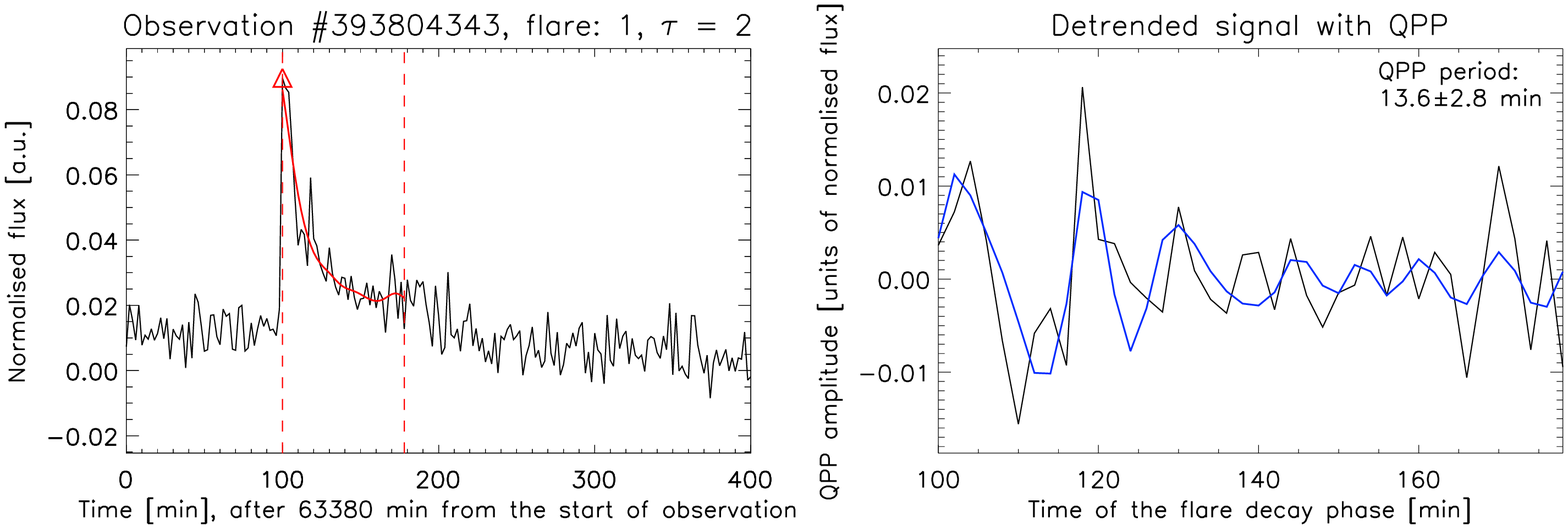}
	\includegraphics[width=\linewidth]{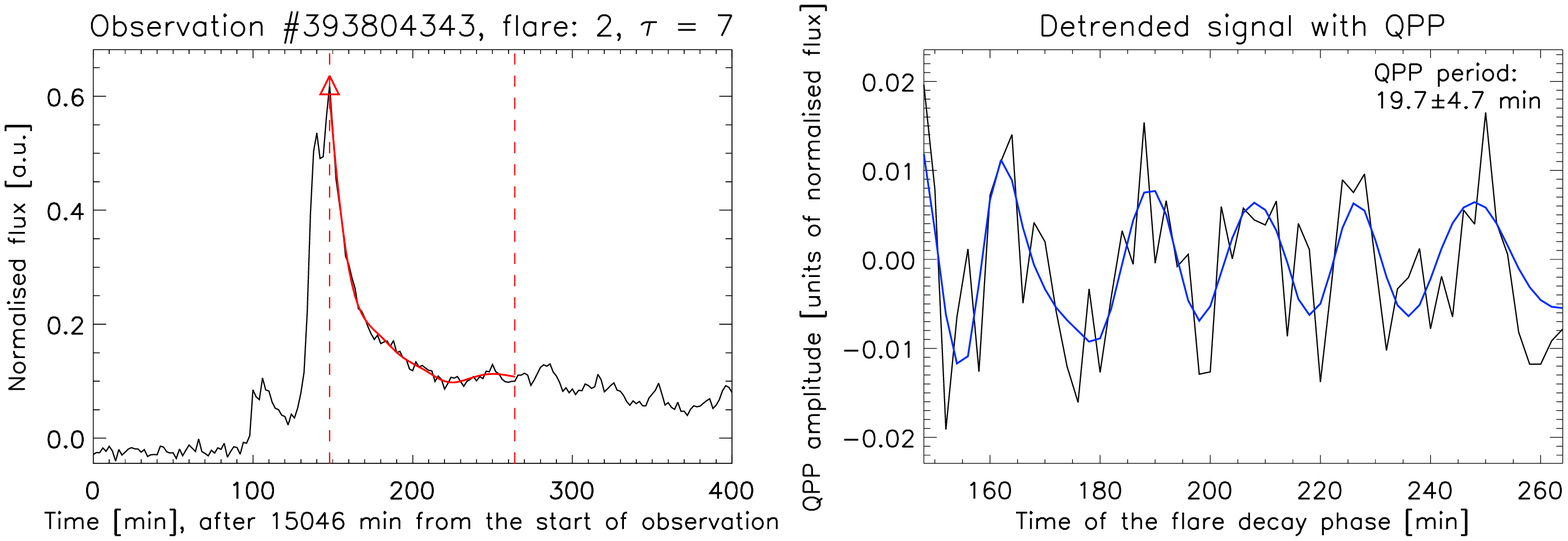}
	\includegraphics[width=\linewidth]{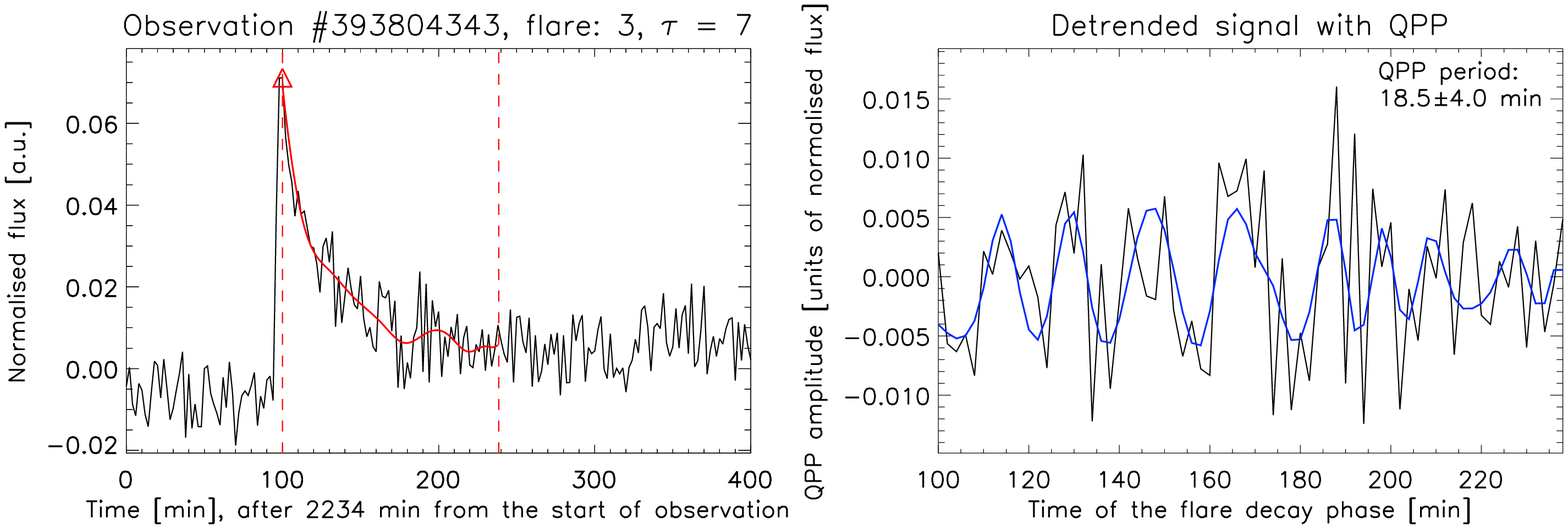}
	\caption{The QPP signals detected in observation TIC 393804343. See Fig.~\ref{fig:qpp_drift} for details on the figure notations.
	}
	\label{fig:qpp_all_3}
\end{figure*}

\end{article} 

\end{document}